\documentclass[prb,aps,amssymb,amsmath,reprint,superscriptaddress,showpacs,floatfix,longbibliography]{revtex4-2}

\newcommand{\mr}[1]{\mathrm{#1}}
\newcommand{\pd}{{\phantom{\dagger}}}

\usepackage{bm}
\usepackage{graphicx}
\usepackage{color}
\usepackage[b]{esvect} % Vector notation 
\usepackage{mathtools}
\usepackage{braket}
\usepackage{amsmath}
\usepackage{amsfonts}
\usepackage{amssymb}
\usepackage{textcomp} % text companion fonts
\usepackage{microtype} 
\usepackage[normalem]{ulem} %to strike the words
\usepackage{hyperref}
\usepackage{float} 
\usepackage{cases}
\usepackage{fancyhdr}

\hypersetup{colorlinks=true}
\hypersetup{citecolor=blue}
\usepackage[all]{hypcap} % let hyperlinks correctly point to figures rather than their captions; 
\begin{document}

\title{Few-body nature of Kondo correlated ground states}

\author{Maxime Debertolis}
\affiliation{Institut N\'{e}el, CNRS and Universit\'e Grenoble Alpes, F-38042 Grenoble, France}
\author{Serge Florens}
\affiliation{Institut N\'{e}el, CNRS and Universit\'e Grenoble Alpes, F-38042 Grenoble, France}
\author{Izak Snyman}
\affiliation{Mandelstam Institute for Theoretical Physics, School of Physics, University of the 
Witwatersrand, Johannesburg, South Africa}

\begin{abstract}
The quenching of degenerate impurity states in metals generally induces a long-range correlated 
quantum state known as the Kondo screening cloud. While a macroscopic number of particles clearly take 
part in forming this extended structure, assessing the number of truly entangled degrees of freedom 
requires a careful analysis of the relevant many-body wavefunction. For this purpose, we examine
the natural single-particle orbitals that are eigenstates of the single-particle density (correlation) 
matrix for the ground state of two quantum impurity problems: the interacting resonant level model (IRLM)
and the single impurity Anderson model (SIAM).
As a simple and general probe for few-body versus many-body character we consider the rate of exponential decay 
of the correlation matrix eigenvalues towards inactive (fully empty or filled) orbitals.
We find that this rate remains large in the physically most relevant region of parameter space, implying a 
few-body character. Genuine many-body correlations emerge only when the Kondo 
temperature becomes exponentially small, for instance near a quantum critical point.
In addition, we demonstrate that a simple numerical diagonalization of the few-body problem
restricted to the Fock space of the most correlated orbitals converges exponentially
fast with respect to the number of orbitals, to the true ground state of the IRLM. 
We also show that finite size effects drastically affect the correlation spectrum, shedding 
light on an apparent paradox arising from previous studies on short chains.

\end{abstract}

\maketitle

\section{Introduction}
\label{s1}

Strongly interacting quantum many-body systems constitute one of the most
challenging problems in physics. The combination of a macroscopic number of
particles with interactions that are relevant in the renormalization sense puts
paid to strategies involving the most commonly used tools of quantum mechanics
(perturbation theory, exact diagonalization). Over the past decades, advanced
numerical methods have been tailored to reliably extract physical information of
interacting fermion models, from the Numerical Renormalization Group
(NRG)~\cite{Wilson_1980} and Density Matrix Renormalization
Group~\cite{White_DMRG} in low dimensions, to continuous time quantum Monte
Carlo simulations~\cite{Gull} within the Dynamical Mean Field
Theory~\cite{Georges1996} for higher connectivity lattices. While answering many
physical questions, these methods have not yet fully characterized the link
between strong correlations and physical complexity for generic quantum many-body
systems.

Some classes of problems that are insurmountable by brute force can
be tackled due to a hidden simplicity of the physically relevant states 
(e.g. ground and low-lying thermal states).
Indeed, diagnostic tools such as entanglement
measures~\cite{Entanglement_Review} have shown that the Density Matrix
Renormalization
group~\cite{OstlundRommer,Dukelsky1998,Vidal2003,Schollwock_Review} owes its
success to the matrix-product state structure of ground states of locally
interacting one dimensional lattices. Insights about entanglement between
spatially distinct regions have subsequently lead to a deep understanding of the
matrix and tensor product state structure of translationally invariant
low-dimensional interacting ground states. The conceptual
understanding of inhomogeneous
systems is less complete, as is attested to by the ongoing work on many-body
localization.~\cite{MBL_Review} A simple starting point for studying
non-uniform many-body states is the class of systems known as quantum impurity
models~\cite{Hewson1993,Bulla2008}: while strong interactions are limited to few
local sites, scattering from electronic reservoirs generates complex quantum
states showing long-range spatial entanglement, dubbed the Kondo screening
cloud~\cite{Affleck1996}. The question of quantifying the amount of
correlations contained in such a non-local many-body impurity state has not yet
been addressed exhaustively~\cite{Park2013,Barcza2020,Borzenets2020}. In this Article, we answer the question
``How many of the particles in the Kondo cloud are correlated with each other or with the impurity in the 
quantum many-body sense?'' 

Given the central position that the Kondo problem occupies in many-body physics, it may
seem the answer is obvious: many electrons become correlated. 
After all, the Kondo screening cloud is typically much larger than the
Fermi wavelength and thus encompasses many conduction electrons.
However, recent studies have come to a different and seemingly paradoxical
conclusion~\cite{Yang2017,Zheng2020}.
These works
considered the one-body density matrix (also called the correlation
matrix)~\cite{He2014,Lu2014,White2015} of the Kondo problem. Its
eigenvectors define an optimal set of single-particle orbitals that are
commonly referred to as ``natural orbitals'' in the quantum chemistry
literature~\cite{CASSCF}. The associated eigenvalues are ground state
occupation numbers for the natural orbitals. If an occupation number is close to
zero or one, i.e. nearly empty of filled, the corresponding orbital is not
involved in many-body correlations, and is therefore said to be inactive. The
remaining orbitals are called active and host correlated particles. One
study~\cite{Zheng2020}
found that there is a single active orbital that is ``solely responsible for screening the impurity spin
in both the weak and strong Kondo coupling regime'', and that the resulting singlet is disentangled from
the rest of the system. The authors of another study~\cite{Yang2017}
similarly report that they have identified
``a dominant single particle wave function that is entangled to the impurity forming a singlet that is, 
to a great extent, practically disentangled from the rest of the conduction electrons''.
We will show that this proposed single-correlated-orbital picture
at weak coupling is purely a finite size effect (here weak coupling means
small exchange interaction, so that the Kondo temperature is exponentially low).
Indeed, in the works quoted, systems consisting of at most a few hundred real-space
lattice sites were studied. However, the Kondo length becomes quickly
larger than this system size when the dimensionless Kondo coupling is 
reduced to values below unity. When this happens, Kondo correlations cannot 
fully develop, and many-body effects are dramatically reduced compared to the thermodynamic 
limit. Clearly, a systematic characterization of the active space of
Kondo-correlated systems in the thermodynamic limit is still lacking, and this
will be one important goal of our study. We point out that the question we are asking concerns 
how to express microscopically the ground state of the system
in terms of the complete set of {\em bare} degrees of freedom, used to define the model. 
It is of course well-known that the {\em effective} description of excitations 
at energies sufficiently smaller than the Kondo temperature, is that of a Fermi liquid.

We have devised the following method to determine the number of particles 
taking part in ground state correlations. 
We use the NRG 
to calculate the correlation matrix of a fermionic quantum impurity model. 
We are particularly interested models that display
Kondo correlations.
The Wilson grid discretization \cite{Bulla2008} employed by NRG allows us to
study systems with a real-space size that grows exponentially with the dimension
of the single-particle Hilbert space. We are thus able to obtain results for the
correlation matrix that are converged to the thermodynamic limit.
We then use the eigenorbitals of the correlation matrix to construct a 
trial state containing $M$ active orbitals on top of an uncorrelated Fermi sea. At half-filling, 
minimizing the energy expectation value of the trial state is equivalent to 
exactly diagonalizing an $M/2$-particle problem in the subspace of active orbitals.
The resulting variational energy is
compared to the true ground state energy (very accurately calculated with NRG).
If the energy difference is much less than the Kondo temperature, then the trial state 
is an accurate approximation to the true ground state. 
When this is the case, we conclude that at most $M$ orbitals ($M/2$ particles) take 
part in correlations. 

We have carried out the above procedure for the interacting resonant level model (IRLM)
that displays {\it bona fide} Kondo correlations in its charge sector.
What we find is surprising: while the picture of a single orbital screening the 
impurity (advocated in Refs.\onlinecite{Yang2017,Zheng2020}) does not apply in general, neither 
does the pessimistic view that Kondo correlations involve a macroscopic number of electrons 
within the large screening cloud. For a realistic Kondo temperature of $10^{-3}$ 
of the ultraviolet scale set by the Fermi energy, we find that a trial state with only 7 
correlated particles approximates the ground state energy to an accuracy of 1\% of the 
Kondo temperature. Only when one reaches unrealistic regimes where the Kondo temperature 
becomes exponentially small, does the number of correlated particles in the ground state 
increase beyond a handful, making unpractical a description in terms of natural orbitals. 
This observation suggests that Fermi liquid ground states of quantum impurity models in the 
thermodynamic limit are for practical purposes few-body in nature, thus neither single-body 
nor many-body, once reformulated in the optimal space of natural orbitals.

An important question concerns whether these results are a general feature of Kondo physics, or 
specific to the IRLM. Arguably, mapping the IRLM to the anisotropic Kondo model 
might yield a more correlated state than the IRLM ground state. Indeed since IRLM fermions are 
non-linear and non-polynomial functions of the bare fermions of the Kondo model,
a few-body correlated IRLM ground state might translate into a truly many-body correlated
ground state in the Kondo representation. For instance, the Toulouse point of
the Kondo model is certainly non-trivial in the original Kondo framework, while it
involves completely free fermions on the IRLM side. We have therefore also studied
the Anderson impurity model (SIAM), that displays Kondo correlations in its spin sector.Our study of the SIAM again reveals an exponential decay of the natural orbitals
to full occupancy or vacancy for any finite Kondo temperature. Our conclusions are therefore not 
specific to the IRLM, and pertain to other quantum impurity models displaying a Fermi liquid 
ground state.

The rest of this Article is structured as follows. In Section \ref{s2} we
introduce the IRLM, and discuss its equivalence to the single channel Kondo model.
We also review general properties of the correlation matrix for quantum impurity
models.
In Section \ref{s3}, we examine numerical results for the correlation matrix spectrum,
using systematic NRG calculations. Special attention is paid to finite size
effects (extra technical details are given in several appendices).
In Section \ref{s4}, we propose a few-body Ansatz based on natural orbitals,
which shows exponential convergence to the numerically exact multi-particle wavefunction
describing our NRG results. Section \ref{s5} contains our results for the correlation matrix 
of the SIAM, which shows that our conclusions are not specific to the IRLM.
Section \ref{s5} summarizes our main findings and identifies promising directions for
future research.

\section{Generalities on the correlation matrix}
\label{s2}

A simple setting to probe Kondo correlated states is the interacting resonant 
level model~\cite{Vigman1978} (IRLM):
\begin{eqnarray}
\nonumber
\cal{H}& = &U\left(d^\dagger d -\frac{1}{2}\right)\left(c_0^\dagger c_0^\pd -\frac{1}{2}\right)
+V\left(d^\dagger c_0^\pd+c_0^\dagger d\right)\\
\label{model}
& &+\sum_{i=1}^{N-2} t_i\left(c_i^\dagger c_{i-1}^\pd+c_{i-1}^\dagger c_i^\pd\right),
\end{eqnarray} 
involving spinless fermions on a tight binding chain of $N$ sites
(including the $d$-level as site $i=-1$). Both Coulomb interaction $U$ and tunneling $V$
couple the resonant level $d^\dagger$ to the local orbital $c^\dagger_0$ at the start of the chain.
Despite the absence of spin degrees of freedom, the IRLM can be mapped onto the 
spin-anisotropic Kondo model~\cite{GiamarchiBook,GogolinBook,WeissBook}.
The mapping is exact for energy scales below the ultraviolet scale set by the Fermi energy measured 
from the bottom of the band, provided the Kondo coupling times the density of states is sufficiently small, 
i.e. one is in the universal Kondo regime.
The equivalence relies on spin-charge separation in the Kondo model, with only spin-density fluctuations
coupling to the magnetic impurity. This subsystem is then bozonized and refermionized in terms of
spinless fermions. The procedure was first outlined in \cite{Guinea}. For a recent review, including
careful bookkeeping of phases generated by fermion exchange, see for instance \cite{Zarand00}.
This equivalence has been used in the past to study some
delicate facets of the Kondo problem with high accuracy, for instance quench
dynamics~\cite{Nghiem2016}.  In the present context, the spinless nature of the IRLM facilitates the
bookkeeping that is necessary to compute accurately the correlation matrix. 

To introduce the correlation matrix, it is helpful to first consider the properties of
uncorrelated fermionic states. These can be viewed as single
Slater determinants, characterized by a set of one-particle orbitals $q_n^\dagger$ that are each either filled or empty.
The orbitals are linear combinations of the physical orbitals $c_i^\dagger$ 
used to construct the Hamiltonian, e.g, the lattice site basis, 
$q_n^\dagger = \sum_i U_{ni} c_i^\dagger$. 
Introducing the correlation matrix 
matrix of the physical orbitals 
$Q_{ij} = \big< c_i^\dagger c^\pd_j \big>$, it is clear that for a Slater determinant
one obtains $\sum_{ij} U_{mi}^{\phantom{*}} U_{ni}^* Q_{ij} = \big< q_m^\dagger
q^\pd_n \big> = \lambda_n \delta_{n,m} $
with $\lambda_n = 0$ or $1$, depending whether orbital $q^\dagger_n$ is empty or filled.
Note that $\hat{Q}$ is proportional to the one-particle reduced density matrix.
Recent studies have used the correlation matrix as a tool to study quantum impurity problems ~\cite{He2014,Lu2014,Zheng2020}.
For a general many-body state, the eigenvalues $\lambda_n$ of the correlation matrix $\hat{Q}$ 
define occupancies between zero and one.
The number of eigenvalues significantly different from zero or one provides a
sensitive measure of correlations, while
the associated eigenvectors of $\hat{Q}$ define the single-particle basis in which 
correlations are most economically represented~\cite{He2014}.
It is worth pointing out that many NRG studies of impurity models
focus on observables associated with the impurity degree of freedom only. 
The $N\times N$ matrix elements $Q_{i,j}=\big<c_i^\dagger c^\pd_{j}\big>$ (where
$i,\,j \in \{-1,\,0,\,\ldots,\,N-2\}$ and $c^\pd_{-1}\equiv d$)
involve observables in the environment ($i,j>-1$) and observables that are hybridized between the
impurity and the environment ($i=-1,j>-1$, $j=-1,i>-1$). Calculating $\hat{Q}$ using NRG is therefore
more involved than the standard NRG analysis of impurity problems.

Let us focus now on the general properties of the correlation matrix $\hat{Q}$.
Clearly, its eigenvalues $\lambda_n$ are independent of the choice of one-particle orbitals
used in its definition.
Since the eigenvalues $\lambda_{n} = \langle q^{\dagger}_{n}q^{}_{n}\rangle$
correspond to occupancies of the natural orbitals, they belong to the interval $[0,1]$.
As mentioned before, the eigenvalues are either zero or one for a Slater determinant, 
and their departure from these trivial values signal that the associated
orbitals participate in quantum many-body correlations.
As a simple example, consider the Bell-like state:
\begin{equation}
\left|\Psi\right>=\frac{1}{\sqrt{2}}\left(c_i^\dagger c_j^\dagger+c_k^\dagger c_l^\dagger\right) \left|\Phi\right>
\end{equation}
with $\left|\Phi\right>$ a Slater determinant that does not involve orbitals $i$, $j$,
$k$, and $l$ (considered distinct from each other).
It is easy to check that for this state $\hat{Q}$ has four eigenvalues different
from zero or one, that are all equal to $1/2$. 
The IRLM Hamiltonian~(\ref{model}) manifests particle-hole symmetry
${\cal{H}}=P^{\dagger} H P$, where $P$ is the unitary and hermitian
particle-hole conjugation operator:
\begin{equation}
P=\prod_{i=0}^{\frac{N}{2}-1}
\left(c_{2i-1}^\pd-c_{2i-1}^\dagger\right)\left(c_{2i}^\pd+c_{2i}^\dagger\right), 
\end{equation}
acting as $P^\dagger c^{}_i P=(-1)^i c_i^\dagger$ and
$P\left|0\right>= c_{-1}^\dagger c_0^\dagger \ldots c_{N-2}^\dagger \left|0\right>$.
Since $P^2=1$, the eigenvalues of $P$ are $\pm1$, and we have
$Q_{ij} = \delta_{ij}-(-1)^{i+j} Q_{ji}$,
thus the diagonal entries of $\hat{Q}$ are all equal to $1/2$.
Furthermore the matrix elements of the Hamiltonian $\cal{H}$ are all real in the
Fock-space basis built from $c_i^\dagger$ operators and hence the expansion
coefficients of the eigenstates of ${\cal{H}}$ in this basis are real too.
This implies $Q_{ij}=Q_{ji}$, and from the particle-hole symmetry of $\hat{Q}$, we 
conclude that $Q_{ij}=0$ for $i+j$ even and $i\neq j$.
Owing to particle-hole symmetry the eigenvalues
of $\hat{Q}$ then come in pairs $1/2\pm r$.

\begin{figure}[htb]
\includegraphics[width=0.99\columnwidth]{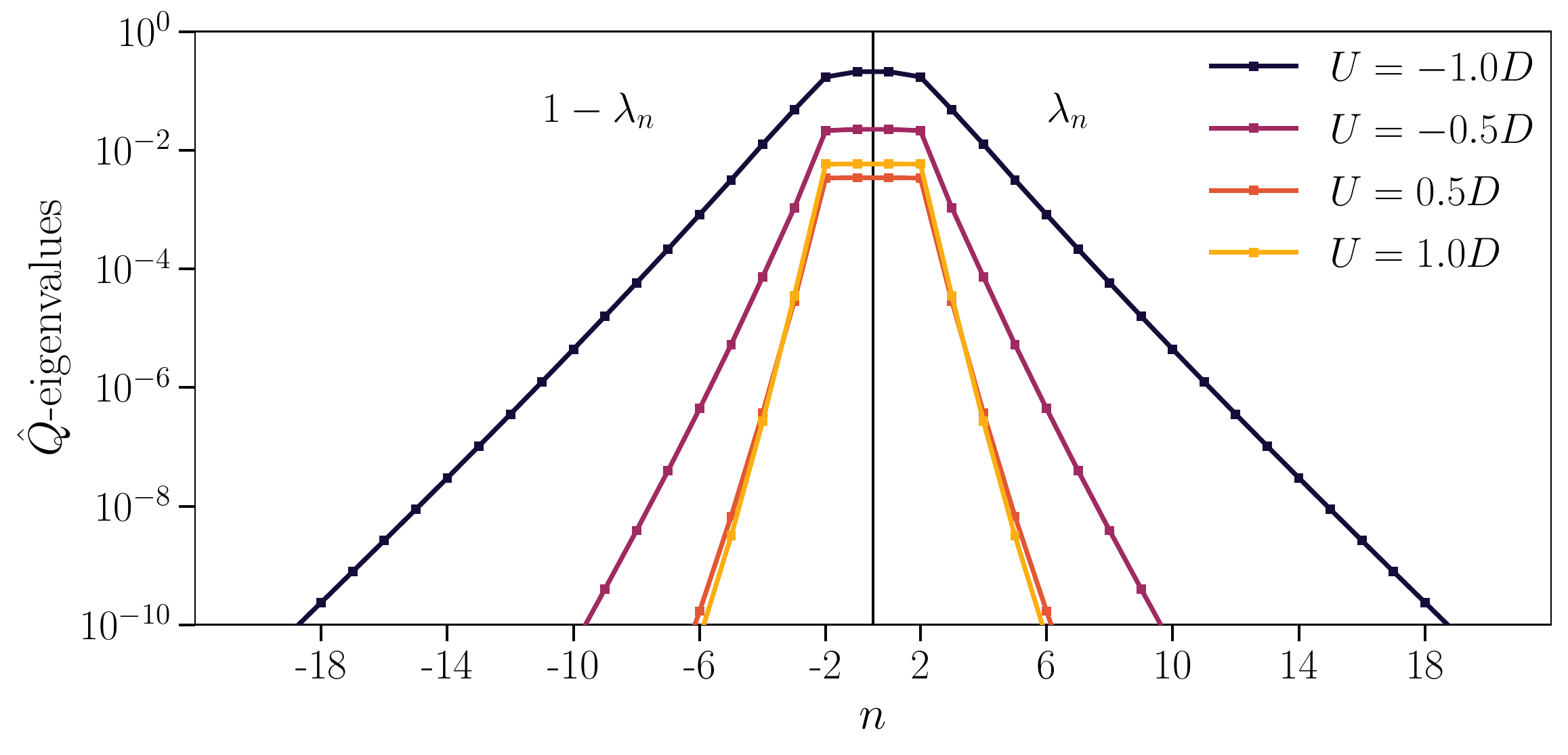}
\caption{Top panel: Spectrum of $\hat{Q}$ for the IRLM at various values of interaction $U$
for $V=0.15$, in units of the half bandwidth $D=1$.
The right side shows eigenvalues $0<\lambda_n < 1/2$, while the left side shows
$1-\lambda_n$ for $1/2<\lambda_n<1$, thus exhibiting particle-hole symmetry
explicitly. Apart from four strongly correlated orbitals on the plateau
(around which the horizontal index $n$ is centered), the rest of the eigenvalues 
decay exponentially fast towards either the empty and filled occupancies. In Appendix \ref{appa}
it is demonstrated that these results are converged to the continuum and thermodynamic limit 
$N\to\infty,\,\Lambda\to1$.
\label{f_qev}}
\end{figure}

\section{Study of the IRLM correlation spectrum}
\label{s3}

The relatively low computational cost to implement NRG for the IRLM makes it possible to track 
with high accuracy the flow of the $N^2$ operators $c^\dagger_i c^\pd_j$ needed for calculating 
$\hat{Q}$ with modest computational resources, provided the block-diagonal structure imposed by 
particle-number conservation is exploited to keep matrix dimensions manageable.
Aiming to resolve $\hat{Q}$-eigenvalues that are exponentially small, 
we calculate the elements of $\hat Q$ to a very high precision, allowing
up to thousands of kept states after truncation (see Appendix \ref{appa} for a detailed study 
of the convergence).
The NRG implementation is based on the hopping amplitudes along the 
Wilson chain~\cite{Bulla2008}:
\begin{equation}
t_j=\frac{\left(1+\Lambda^{-1}\right)\left(1-\Lambda^{-j-1}\right)}
{2\sqrt{1-\Lambda^{-2j-1}}\sqrt{1-\Lambda^{-2j-3}}}\Lambda^{-j/2} D,\label{deft}
\end{equation}
so that $D=1$ sets the half-bandwidth of the bath and also our Fermi energy.
We present here calculations for the Wilson parameter $\Lambda=1.5$ (a more systematic 
study is presented in Appendix \ref{appa}), tunneling $V=0.15\,D$, and up to
$N=180$ sites. The lowest energy at play is thus of
the order $\Lambda^{-N/2}D\simeq 10^{-16}\,D$, ensuring convergence to the
ground state for all practical purposes.

Fig.~\ref{f_qev} displays the full eigenspectrum of $\hat{Q}$, showing on the left
side $1-\lambda_n$ for $1/2<\lambda<1$, and on the right side $\lambda_n$ for $0<\lambda_n<1/2$, so
that particle-hole symmetry becomes apparent. Note that the eigenvalue index $n$ runs from
$-N/2$ to $N/2$, excluding $n=0$, in order to display more clearly the particle-hole conjugation.
The general behavior of the particle-hole 
symmetrized spectrum is as follows. There is an approximate four-fold degeneracy of the 
highest eigenvalue $\lambda_\mr{max}\equiv\lambda_1=1-\lambda_{-1}$, indicating Bell-like entanglement between the 
four most correlated orbitals $q^\dagger_{-1}$, $q^\dagger_{-2}$, $q^\dagger_{1}$, 
$q^\dagger_{2}$ (from our chosen convention, the index $n$ is centered around those
most correlated orbitals).
This is related to the fact that, at negative $U$, the impurity orbital and the first energy 
shell tend to be either both filled or both empty due to Coulomb attraction, and similarly, at
positive $U$, if the impurity orbital is filled, the first energy shell tends to
be empty, and vice versa. 
The other eigenvalues decay exponentially, $\lambda_n\simeq A e^{-x n}$ for
$n>2$ and $\lambda_n\simeq 1- A e^{+x n}$ for $n<-2$, with a decay rate $x$ that depends 
on interaction strength. We show in Appendix \ref{appa} that this
exponential decay is not an artefact of the Wilson chain, and is robust in
the continuum limit $\Lambda\to1$. This behavior of the $\hat{Q}$-eigenvalues has previously been observed in
studies of impurity models discretized on regular real space latices~\cite{He2014,Lu2014,Zheng2020}.
We emphasize that the Kondo regime corresponds to $U<0$ in the IRLM,
and indeed the slower decay of the $\hat{Q}$-eigenvalues in
Fig.~\ref{f_qev} attests that this regime is more correlated than for $U>0$.

\begin{figure}[ht]
\begin{center}
\includegraphics[width=.90 \columnwidth]{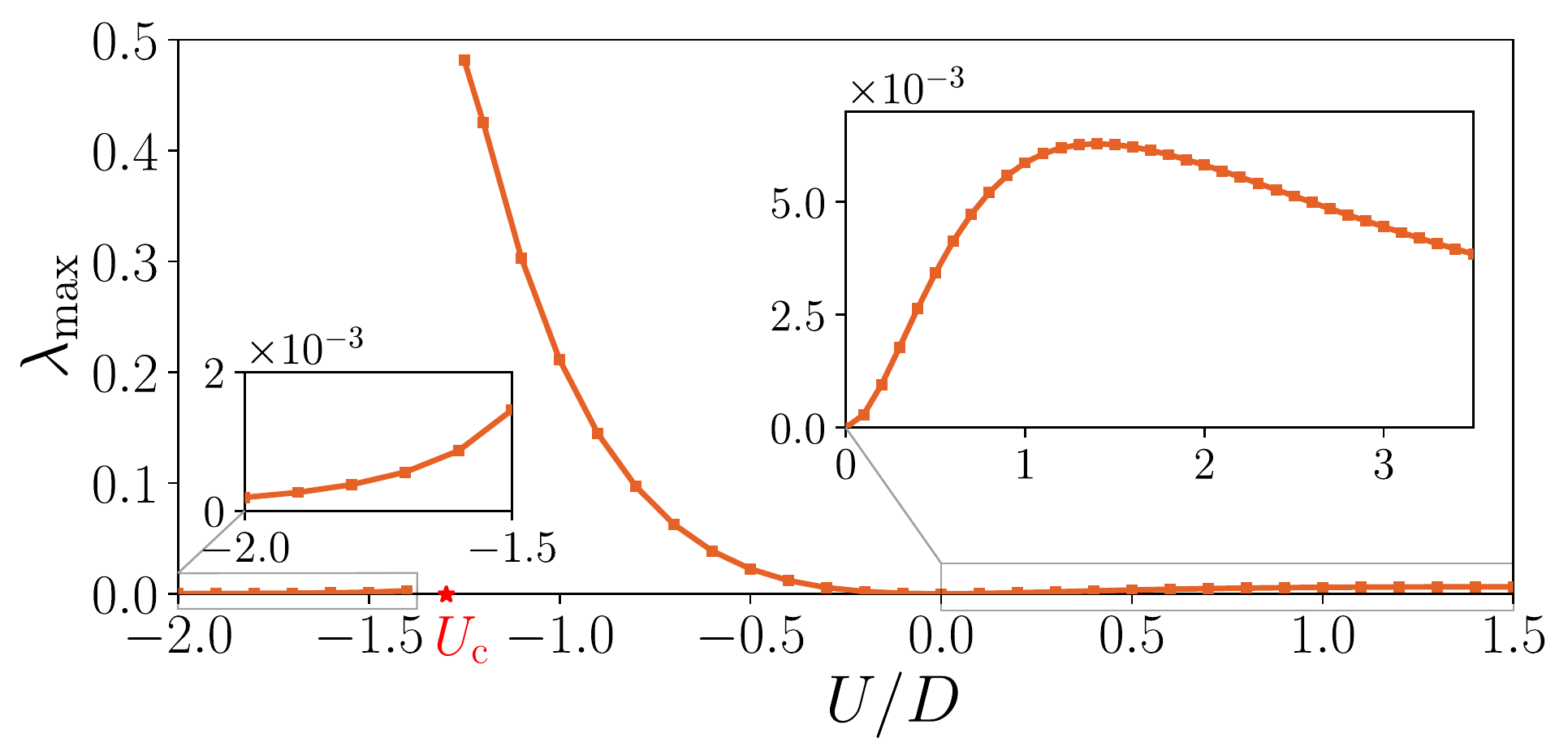}
\phantom{sssss}
\includegraphics[width=.99 \columnwidth]{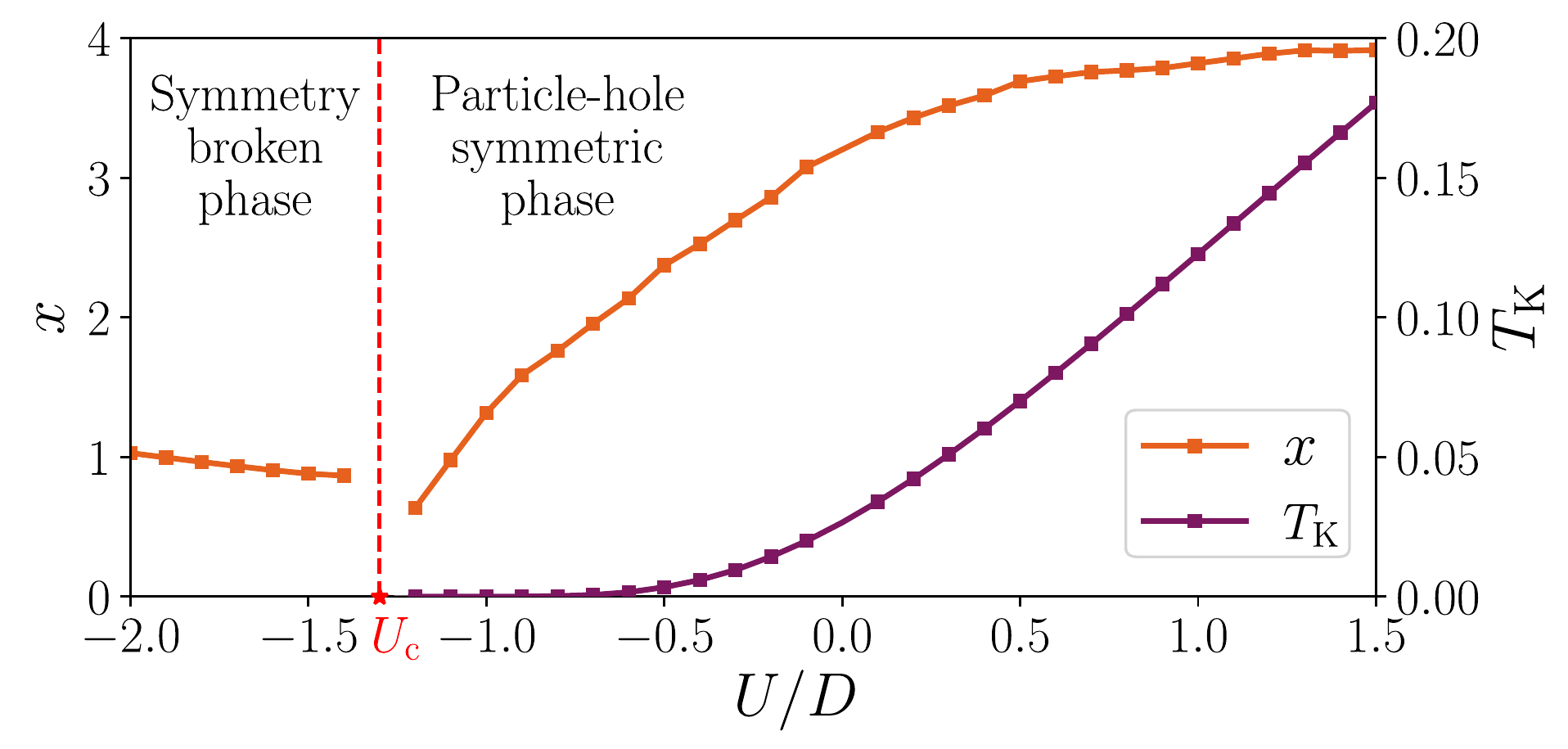}
\phantom{s}\hspace{-1.2cm}
\includegraphics[width=.90 \columnwidth]{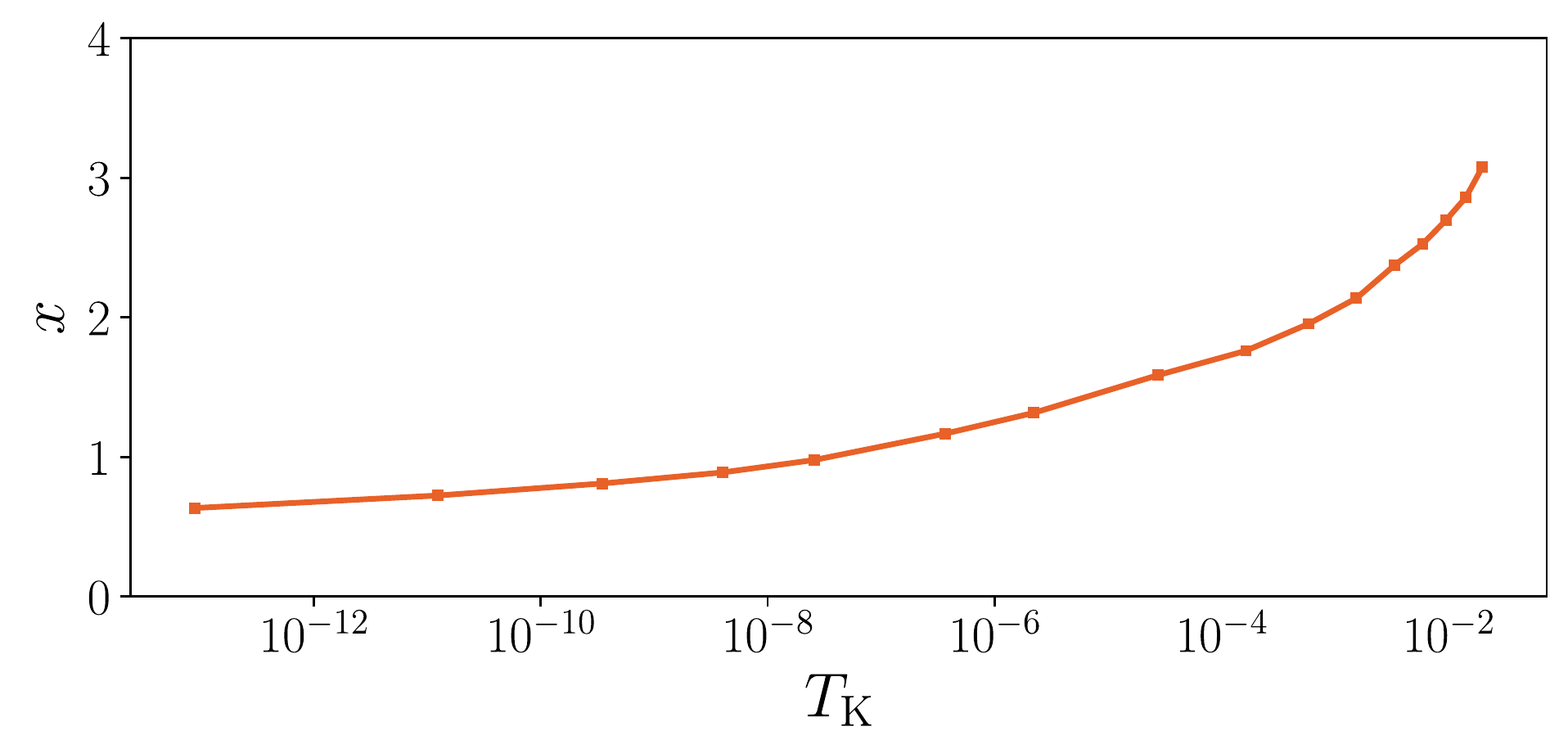}
\end{center}
\caption{Top panel: Maximum $\hat{Q}$-eigenvalue $\lambda_\mr{max}$ as a 
function of interaction $U$. Middle panel: Decay rate $x$ of the $\hat{Q}$-eigenvalue
(upper curve) and Kondo temperature $T_K$ (lower curve) versus $U$.
Only the quantum critical regime near $U_c\simeq -1.3$ is strongly correlated, 
since the $Q$-eigenvalues show both a slow decay ($x<1$) and an enhanced 
$\lambda_\mr{max}\simeq0.5$.
Bottom panel: Decay rate $x$ replotted as a function of Kondo temperature $T_k$, showing 
a slow inverse logarithmic decrease of $x$ when $T_k$ vanishes at the IRLM quantum critical 
point.} 
\label{LambdaMax}
\end{figure}

The behavior of $\lambda_\mr{max}$, the maximum eigenvalue of $\hat{Q}$ in the
range $[0,1/2]$, is displayed as a function of interaction $U$ in
Fig.~\ref{LambdaMax}, showing that it remains small for all $U>0$ (this is the
weakly correlated sector of the IRLM), vanishes at $U=0$ (the ground state is a
Slater determinant, so that all eigenvalues are trivial), and increases sharply
only for $-1.3<U<-1.0$ due to the approach to the IRLM quantum critical point
$U_c=1.3$ where the Kondo temperature vanishes. Our key observation is
that the decay rate $x$ of the $\hat{Q}$-eigenvalues drops to small values only when the 
Kondo temperature becomes exponentially small, seemingly with a linear vanishing 
as $|U-U_c|$ (see middle panel), as also shown by the slow inverse logarithmic decrease 
of $x$ as a function of Kondo temperature (see bottom panel in Fig.~\ref{LambdaMax}).
Thus, only the quantum critical regime corresponds to a true many-body state
as opposed to a few correlated particles on top of an uncorrelated Fermi sea.
Larger negative $U<-1.3$ leads to a discontinuous transition to a phase where
particle-hole symmetry is broken (corresponding to the ferromagnetic phase 
of the Kondo Hamiltonian), involving clearly less correlations due to a 
jump of the decay rate $x$ to finite values.

\begin{figure}[h]
\begin{center}
\includegraphics[width=.99 \columnwidth]{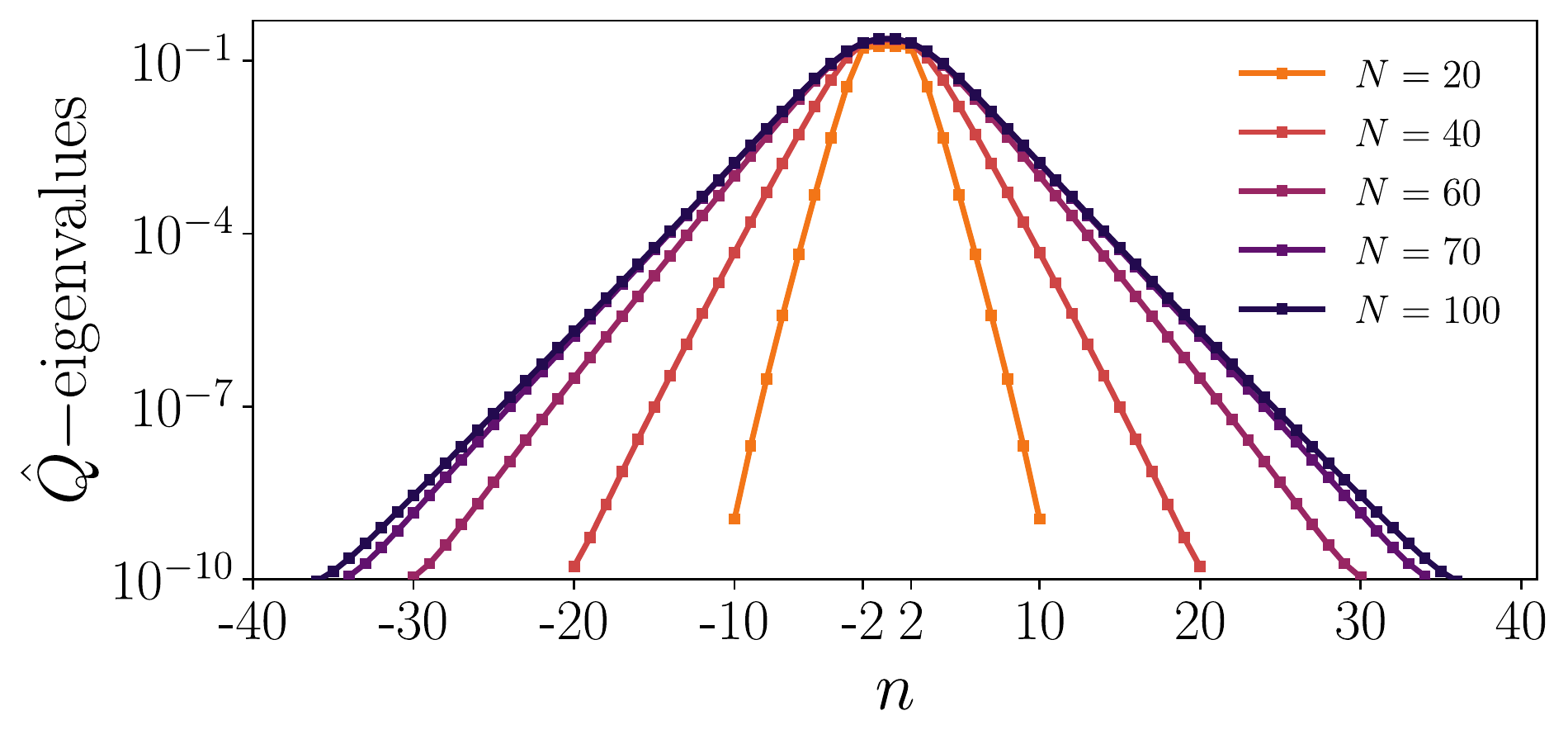}
\end{center}
\caption{Finite size scaling of the $\hat{Q}$-matrix spectrum for $U=-1.2$ near 
the quantum critical point, for various system sizes $N$.
In order to reach the exponentially long Kondo length $1/T_{\rm K}=10^{13}$,
a slightly larger discretization parameter $\Lambda=2.25$ was used.
The corresponding real-space system size $\Lambda^{N/2}$ equals
the Kondo length when $N=75$, as seen by the convergence to the thermodynamic
limit for $N>80$. Smaller systems are plagued by finite size effects that
preempt the full formation of the Kondo state, and show a very rapid decay of the
$\hat{Q}$-eigenvalues.}
\label{ev_vs_size}
\end{figure}

Our results are seemingly inconsistent with results for the Kondo model
reported in ~\cite{Yang2017,Zheng2020}. According to these studies, $x$ should
become large close to the phase transition, whereas we find that it vanishes.
To shed light on the apparent paradox, we show in Fig.~\ref{ev_vs_size} the
spectrum of $\hat{Q}$ as a function of the number $N$ of Wilson chain sites. The
real space system size is $\Lambda^{N/2}$. Here we used $\Lambda=2.25$ and
picked a point $U=-1.2$ close to the critical point. For this choice, the Kondo
length has the astronomically large value $1/T_{\rm K}=10^{13}$, which matches the system 
size when $N=75$.  (In Appendix \ref{appc} we explain how the Kondo temperature $T_{\rm K}$ is
calculated.) For $N$ significantly smaller than $75$, we see very quick
exponential decay of the spectrum, corresponding to large $x$ and a ground state
with few correlated particles. However, when $N$ increases beyond $75$, the
decay rate $x$ soon saturates to a small value, so that a large number
of correlated particles participate in the true ground state in the 
thermodynamic limit. These finite size artifacts explain the results reported
in \cite{Yang2017,Zheng2020}
where real space lattices with at most a few hundred sites were studied, leading
to system sizes of the order of a hundred times the Fermi wavelength. The
exponentially diverging Kondo length reaches this order of magnitude long before
the weak coupling regime in the vicinity of the critical point is entered. In
terms of IRLM parameters, a system size between $10^2$ and $10^3$ times the
Fermi wavelength prevents a fully correlated ground state from forming for
$U<-0.5$, and leads to a severe overestimate of the decay rate $x$ close to the
critical point. Figure~\ref{ev_vs_size} clearly establishes that
correlations increase when the Kondo cloud becomes more extended, in agreement
with intuition.

\begin{figure}[h]
\begin{center}
\includegraphics[width=.99 \columnwidth]{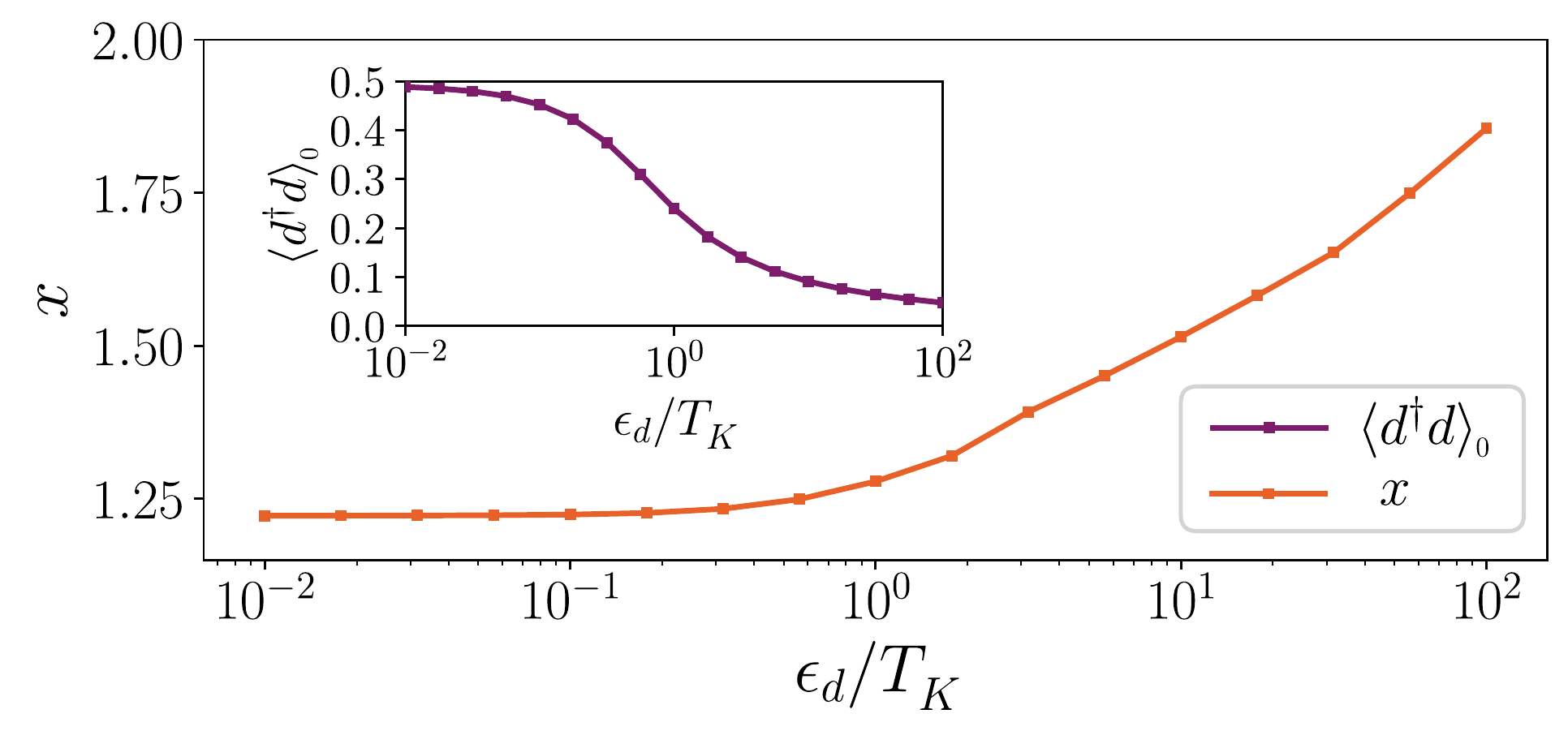}
\end{center}
\caption{
Main panel: Decay rate $x$ versus $d$-level on-site energy $\epsilon_d$, at $U=-1.0$, 
$V=0.15$ and discretization parameter $\Lambda=1.5$.
The corresponding Kondo temperature is $T_{\rm K}=1.87\times10^{-6}$.
Correlations are clearly weakened ($x$ increases) when breaking the charge
degeneracy of the $d$-level as $\epsilon_d$ increases. Inset: ground-state 
d-level occupancy $\left< d^\dagger d\right>$ versus $d$-level on-site energy $\epsilon_d$, 
for the same parameters as main panel.}
\label{x_v_ed}
\end{figure}

A standard method for probing Kondo correlations is to perturb the system at the Kondo scale, and
to see the effect this has on observables. For instance, a biasing potential $\epsilon_d d^\dagger d$ in the IRLM,
corresponding to a Zeeman splitting between the spin-up and spin-down states of the magnetic
impurity in the Kondo model, prevents formation of the Kondo singlet. The occupancy
$\left<d^\dagger d\right>$ (or equivalently the impurity magnetization) reveals significant
symmetry breaking when $\epsilon_d$ reaches the Kondo scale. It is intuitively clear
that the symmetry breaking in the ground state is a sign of reduced correlations, but
the observable $\left<d^\dagger d\right>$ does not directly measure this -- 
one can clearly modify the degree of correlations in the ground state without changing 
$\left<d^\dagger d\right>$ at half-filling ($\epsilon_d=0$) when $U$ is
changed.
The spectrum of $\hat{Q}$, on the other
hand, directly measures correlations. 
In Fig.~\ref{x_v_ed} we plot the decay rate $x$ of the $\hat{Q}$-eigenvalues as a function of $\epsilon_d$. 
The calculation was performed for $U=-1.0$ and $V=0.15$, which corresponds to $T_{\rm K}=1.87\times10^{-6}$.
We used $\Lambda=1.5$ which yields a decay rate $x$ that is converged to the $N\to\infty,\,\Lambda\to1$ limit.
For comparison, we also plot $\left<d^\dagger d\right>$ versus $\epsilon_d$ in an inset.
We see that $x$ starts changing from its unperturbed value when $\epsilon_d$ exceeds the Kondo temperature. 
As $\epsilon_d$ increases further,
$x$ increases monotonically, indicating that fewer and fewer correlated particles are present,
the more severely singlet formation is prevented. In this way, the $\hat{Q}$ matrix spectrum
proves the picture suggested by the $d$-level occupancy $\left<d^\dagger d\right>$.

\begin{figure}[h]
\begin{center}
\includegraphics[width=.99 \columnwidth]{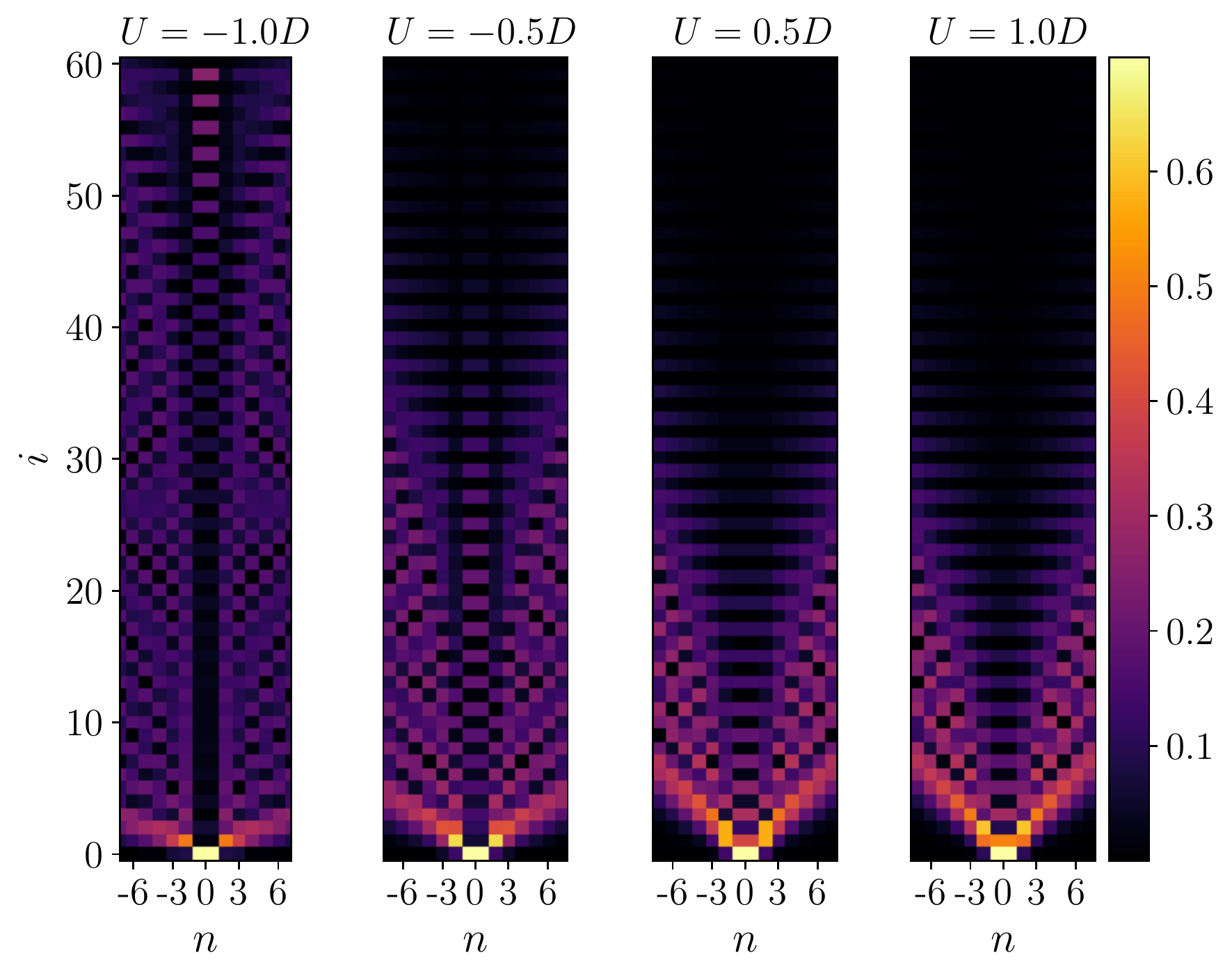}
\end{center}
\caption{Spatial dispersion of the 14 most correlated orbitals $q_n^\dagger$
($n=-7\ldots7$) along the Wilson chain (sites $i=-1\ldots60$), given 
by the absolute value of eigenvectors $|U_{ni}|$. 
The development of Kondo correlations for $U<0$ is evidenced by the long spatial
tails, especially at $U=-1.0$, while little is changed in the spatial profile for 
the weakly correlated regime $U>0$.}
\label{spatial}
\end{figure}
We now examine the spatial dispersion (along the Wilson chain) of the $\hat{Q}$-matrix 
orbitals $q_n^\dagger = \sum_i U_{ni} c_i^\dagger$, by plotting the absolute value 
$|U_{ni}|$ of the eigenvectors obtained from the diagonalization of the matrix $Q_{ij}$ (this
also displays particle-hole symmetry more clearly).
Fig.~\ref{spatial} shows how correlations spread along the system for
four values of the interaction $U$.
It is clear from Fig. \ref{spatial} that all the natural orbitals are highly non-local, 
and carry information mostly forward along the chain. The most correlated orbitals 
($n=1$ and $n=-1$) are predominantly localized near the impurity (site $i=-1)$, as expected 
from the short range of the interaction, but develop also long tails that extend 
to large distances.
For $U>0$, the spatial structure of correlated orbitals is fairly insensitive to the interaction strength, 
showing that this regime remains weakly correlated.
In contrast, for negative values of the interaction, as we go closer to the
quantum critical point $U_c=-1.3$, correlated orbitals become more delocalized, due to the divergence of the Kondo length. 
In addition, more and more orbitals 
become entangled, due to the slower decay of the eigenvalues $\lambda_n$ in Fig.~\ref{f_qev}.
 
\section{Few-body Ansatz from natural orbitals}
\label{s4}
Equipped with this construction of the natural orbitals of the $\hat{Q}$-matrix, we establish 
our most surprising finding, namely that the ground state of the IRLM
is few-body in nature for realistic ({\it i.e.} non exponentially vanishing) Kondo temperatures. 
This result is clearly suggested by the exponential decay of the 
$\hat{Q}$-matrix eigenvalues in Fig.~\ref{f_qev}. Since most of the eigenvalues $\lambda_n$ 
are exponentially close to either zero or one, it seems a good approximation to assume that 
their associated orbitals are exactly uncorrelated, keeping a core of $M$ truly correlated 
orbitals within the ground state wave function (those orbitals correspond to the
$M$ $\lambda_n$-eigenvalues
that are closest to 1/2, and we choose $M$ to be even, which allows for a correlated sector that is exactly half-filled).
Specifically, half of the $N-M$ uncorrelated orbitals (the ones that have 
their eigenvalues closest to $1$) will be frozen and described by a Slater determinant 
$\left|\Psi_0\right>=\prod_{m=-\frac{N}{2}}^{-\frac{M}{2}-1} q^{\dagger}_{m}\left|0\right>$
in the eigenorbitals of the correlation matrix computed by NRG.
The other half of the uncorrelated orbitals (those with $\hat Q$-eigenvalues closest to $0$) 
are taken as empty.
We therefore write the full wave function as follows:
\begin{equation}
|\Psi_\mr{few}\rangle =
{\sum_{\{N_n\}}}
\Psi(N_{-\frac{M}{2}},\ldots, N_{\frac{M}{2}})\!\!\!\!
\prod_{n=-\frac{M}{2}}^{\frac{M}{2}}\!\!
[q^{\dagger}_n]^{N_n}
|\Psi_0\rangle,
\label{FewBodyWF}
\end{equation}
with $N_n=0,1$ the occupancy of correlated orbital $q^\dagger_n$, the summation 
restricted to occupations such that $\sum_{n=-M/2}^{M/2}N_n=M/2$, and
$\Psi(N_{-\frac{M}{2}},\ldots, N_{\frac{M}{2}})$ the complete few-body wave function 
in the correlated subspace.
Note that the total set of $\hat{Q}$-orbitals runs with index $n=-N/2,\ldots,N/2$,
as in Fig.~\ref{f_qev}, and that the index $n=0$ is excluded in the above 
expression. We stress that such an Ansatz is very common in quantum chemistry, where
an active space (also dubbed the correlated sector) is used to select the most 
important chemical degrees of freedom~\cite{CASSCF}.

The Hamiltonian can be re-expressed within the $q_n^\dagger$ orbitals, and then exactly 
divided into three pieces: ${\cal{H}} = {\cal{H}}_\mr{corr} + {\cal{H}}_\mr{uncorr}
+ {\cal{H}}_\mr{mix}$, depending on whether the indices $n$ act only within 
the correlated sector (first term), or only within the uncorrelated sector
(second term), or mix both sectors (third term). 
Minimizing $\langle \Psi_{\rm few}|{\cal{H}}| \Psi_{\rm few}\rangle$
with respect to the few-body wave function $\Psi(N_{-\frac{M}{2}},\ldots, N_{\frac{M}{2}})$
yields a variational energy equal to the ground state energy of the few-body Hamiltonian
${\cal{H}}_\mr{few} = {\cal{H}}_\mr{corr} +
\Pi\left({\cal{H}}_\mr{uncorr}+{\cal{H}}_\mr{mix}\right)\Pi^\dagger$, that acts
on states in which electrons occupy correlated orbitals only, with
$\Pi=\prod_{m=-\frac{N}{2}}^{-\frac{M}{2}-1} q_{m}$.
Within the Fock space constructed from correlated orbitals only,
$\Pi {\cal{H}}_\mr{uncorr}\Pi^\dagger$ is a real number, while 
$\Pi {\cal{H}}_\mr{mix}\Pi^\dagger$ is a quadratic operator
(see Appendix \ref{appb} for details).
The optimal wavefunction $\Psi(N_{-\frac{M}{2}},\ldots, N_{\frac{M}{2}})$ can be found
by exact diagonalization of the few-body Hamiltonian ${\cal{H}}_\mr{few}$, which
we have done for increasing values of $M$.

\begin{figure}[ht]
\begin{center}
\includegraphics[width=.99 \columnwidth]{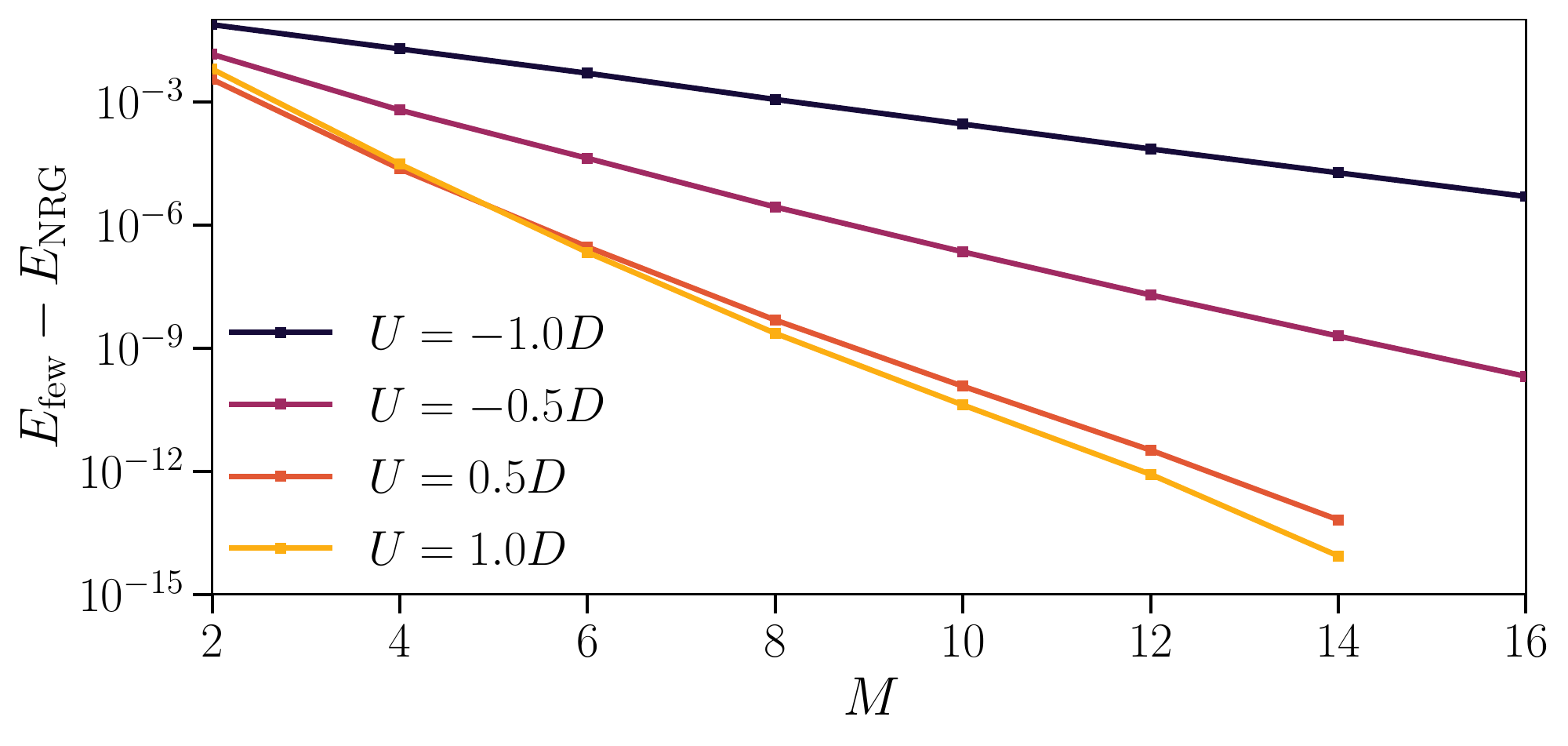}
\end{center}
\caption{Difference between the ground state energy $E_\mr{few}$ computed with 
the few-body wavefunction~(\ref{FewBodyWF}) and the numerically exact energy 
$E_\mr{NRG}$ obtained from NRG, as a function of the number of correlated
orbitals.}
\label{f_conv}
\end{figure}

The only relevant parameter of the few-body approximation is the number $M$ of
kept correlated orbitals. Obviously, the limit $M\to N$ would lead to the exact
wave function.
We stress that all computations at finite $M$ are done in the thermodynamic limit, since 
the $N-M$ uncorrelated orbitals are fully accounted for in our
Ansatz~(\ref{FewBodyWF}). These uncorrelated orbitals actually constitute a
major part of the total energy, despite being evaluated in a single-particle
picture.
The difference between the computed few-body energy $E_\mr{few} = 
\big<\Psi_\mr{few}|{\cal{H}}|\Psi_\mr{few}\big>$ at fixed $M$ and the many-body
ground state energy $E_\mr{NRG}$ obtained from the converged NRG simulations is
shown in Fig.~\ref{f_conv}.
We find an exponential convergence of the few-body energy as a function of the
number $M$ of correlated orbitals, as anticipated from the structure of the $\hat{Q}$-matrix spectrum. 
Note that for the half-filling considered here, the number of truly interacting
fermions is $M/2$. We see that an accuracy of 6 digits is obtained for 6
correlated orbitals (3 interacting particles) for all $U>0$. For $U<0$, the rate of convergence 
becomes slower the closer we come to the critical point, consistent with the increase in the number
of $\hat{Q}$-eigenvalues that are significantly different from zero or one.
However, even at $U=-1.0$, where the Kondo temperature is $1.87\times10^{-6}$,
20 correlated orbitals (10 interacting particles) would give an accuracy better than 10\% of the Kondo temperature.
This clearly vindicates our claim that for practical purposes, the ground state of the Kondo problem is 
 few-body and not many-body in nature, once expressed in the 
optimal set of natural orbitals. Only exponentially close to a quantum phase transition does 
a truly many-correlated-particle wave function emerge, as discussed previously
for dissipative systems~\cite{BeraSubohmic}.
\section{SIAM Correlation spectrum}
\label{ssiam}
In this final section, we investigate whether our results extend to other
impurity models, by investigating the correlation spectrum of the single impurity 
Anderson model (SIAM).
The IRLM and SIAM share the same universal low energy 
physics at scales that are small compared to the ultraviolet cutoff, although 
Kondo correlations pertain to the charge sector of the IRLM and to the spin sector of the SIAM.
Does this automatically mean that the ground state of the SIAM (and other models in
the same Kondo universality class) is few-body in nature, 
as long as $T_K$ is finite but sufficiently smaller than the ultraviolet cutoff?
The universality of Kondo physics does not settle this question, for the following reason. 
The fermionic operators that appear in the IRLM are non-linear (and non-polynomial) functions of 
the ones appearing in the definition of the anisotropic Kondo Hamiltonian. Our results up to this point show that the Kondo ground state is effectively few-body in nature, 
when expressed in terms of IRLM fermions, and it remains to be checked whether this is also
true in other representations, such as the arguably more fundamental fermions of the SIAM.
 
To address this question, we investigate the correlation matrix of the single impurity Anderson
model (SIAM)
\begin{eqnarray}
\nonumber
\cal{H}& = &U\left(d^\dagger_\uparrow d_\uparrow -\frac{1}{2}\right)
\left(d^\dagger_\downarrow d_\downarrow -\frac{1}{2}\right)\nonumber\\
&&+ V \sum_{\sigma=\uparrow,\downarrow}
\left(d^\dagger_\sigma c_{0,\sigma}+c_{0\sigma}^\dagger
d_\sigma\right)\nonumber\\
&&+\sum_{\sigma=\uparrow,\downarrow}\sum_{i=1}^{N-2} t_i\left(c_{i,\sigma}^\dagger
c_{i-1,\sigma}^\pd+c_{i-1,\sigma}^\dagger c_{i,\sigma}^\pd\right),
\label{eq:HSIAM}
\end{eqnarray} 
whose effective low-energy description in the strong interaction limit $U\gg\Gamma=V^2/(2D)$
is the Kondo model, $D$ being the half-bandwidth. Hamiltonian~(\ref{eq:HSIAM})
is again discretized on the Wilson chain, and we used particle number as well as spin conservation
to optimize the numerical simulations, as the spinfulness of the SIAM fermions doubles the dimension of 
the single-particle Hilbert space. Because the ground state is a spin singlet,
$\big<c_{i\sigma}^\dagger
c_{j\sigma'}\big>\propto\delta_{\sigma\sigma'}$ and 
$\big<c_{i\uparrow}^\dagger
c_{j\uparrow}\big>=\big<c_{i\downarrow}^\dagger
c_{j\downarrow}\big>$. Thus there is an extra two-fold degeneracy in the 
correlation matrix spectrum, as compared to the IRLM. The NRG calculation of the correlation matrix
demands more computational resources than for the IRLM, but as we show in Appendix \ref{appa}, we
succeeded in obtaining well-converged results. As for IRLM, we include the $d$-level
fermions in the operators used to construct the correlation matrix, so that the limit $U=0$ is
strictly uncorrelated.
\begin{figure}[htb]
\begin{center}
\includegraphics[width=.99 \columnwidth]{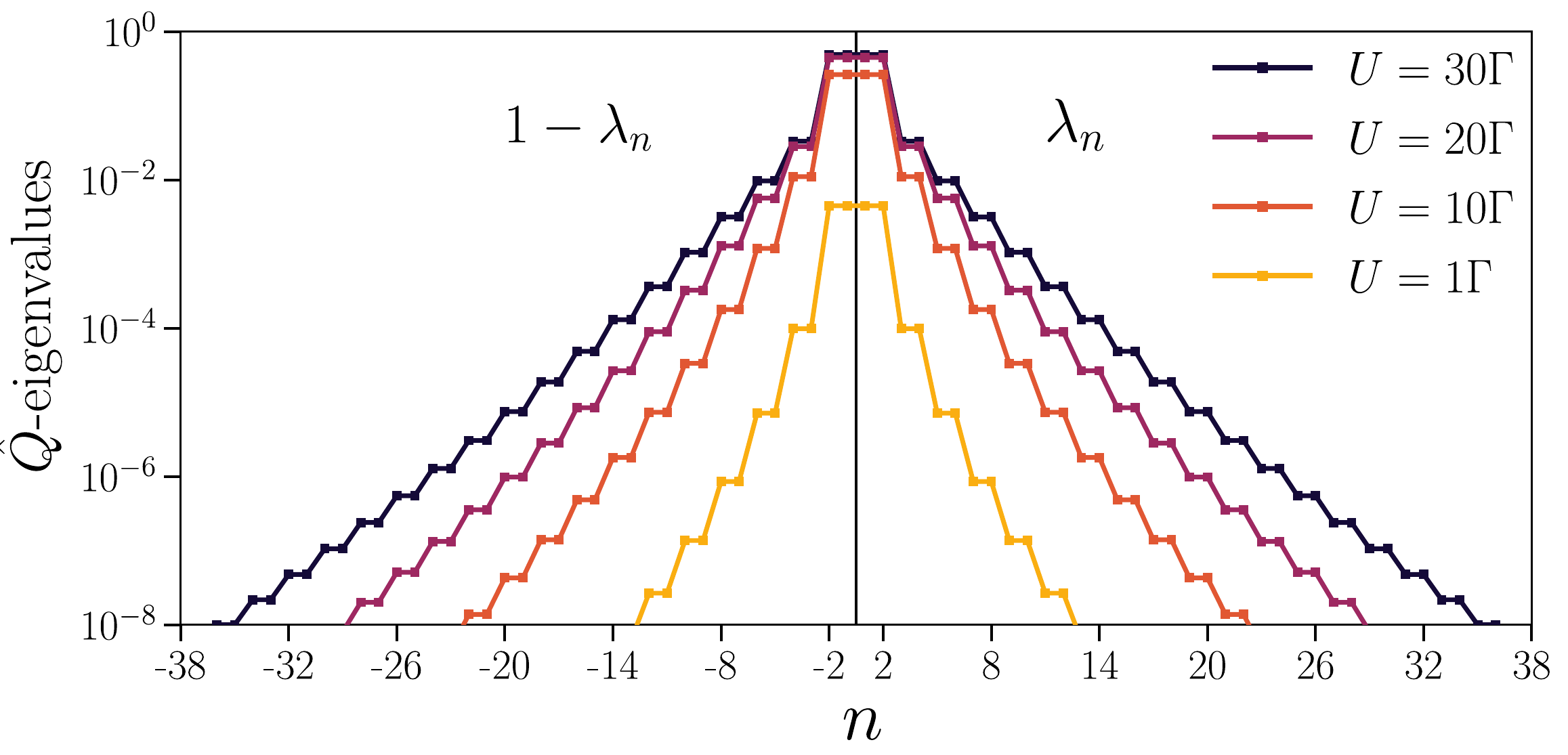}
\includegraphics[width=.99 \columnwidth]{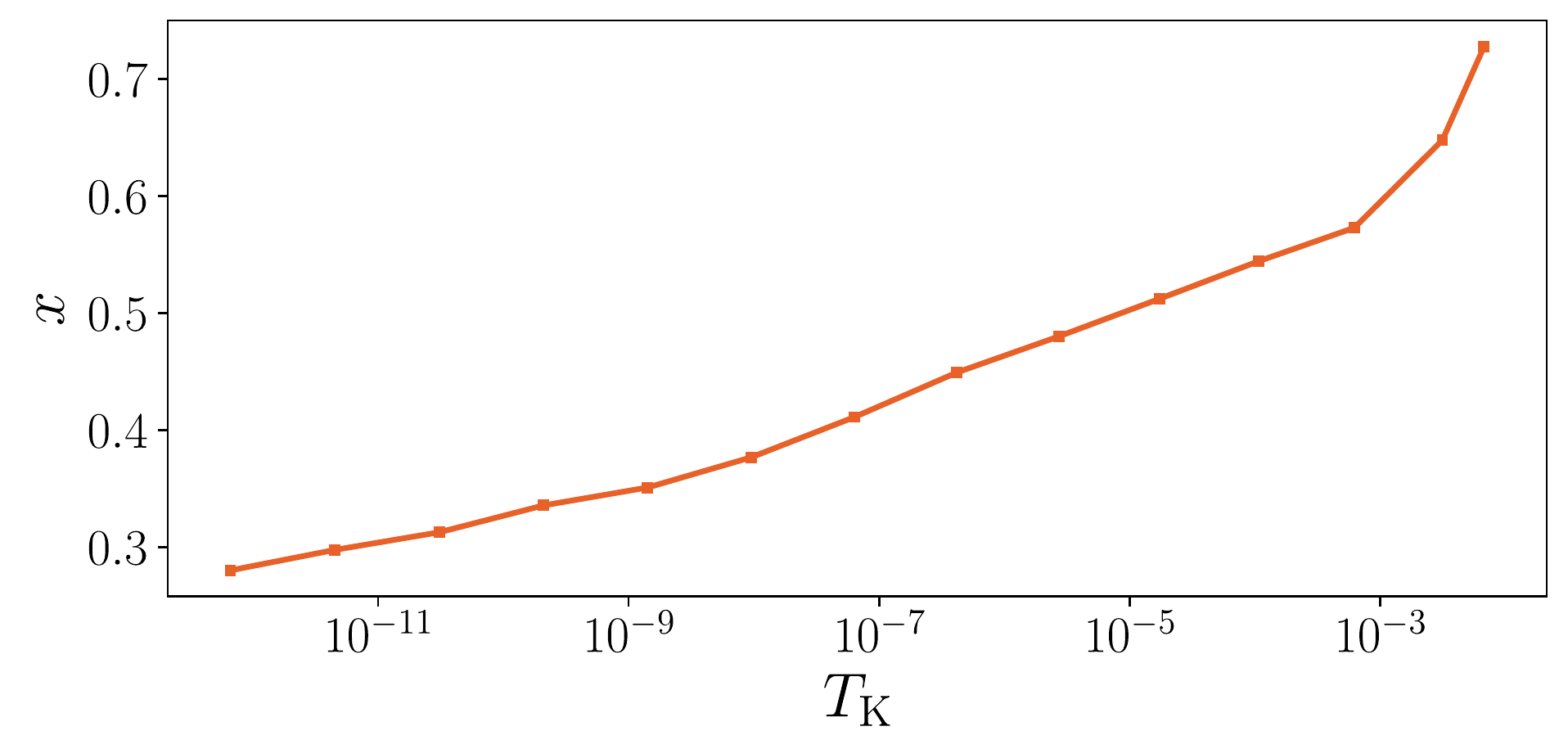}
\end{center}
\caption{Top panel: Correlation matrix spectrum for the single impurity Anderson model (SIAM)
for various ratios of the onsite interaction $U$ to the hybridization $\Gamma$. In all cases 
$\Gamma=0.01D$.
As in Fig.\,\ref{f_qev}, the right
side shows eigenvalues $0<\lambda_{n\sigma} < 1/2$, while the left side shows
$1-\lambda_{n\sigma}$ for $1/2<\lambda_{n\sigma}<1$. Spin degenerate pairs 
of eigenvalues $\lambda_{n\sigma}$ decay exponentially $\sim\exp(-|n|x)$
to full occupancy or vacancy. Bottom panel: Decay rate $x$ of correlation
matrix eigenvalues, versus Kondo temperature $T_k$.
\label{siam}}
\end{figure}

The top panel of Fig.\,\ref{siam} shows the
correlation matrix spectrum, plotted in the same way as for the IRLM in Fig.\,\ref{f_qev}. Beyond a
central plateau, that still contains the four eigenvalues furthest from full occupancy or vacancy,
we see degenerate pairs $\lambda_{n\uparrow}=\lambda_{n\downarrow}$ that decay exponentially
$\lambda_{n\sigma}\sim\exp(-x|n|)$. In the bottom panel of Fig.\,\ref{siam}, we show the extracted
decay rate $x$, as a function of the Kondo temperature $T_k$ (the latter is calculated from the magnetic 
susceptibility of the impurity). We see a finite decay rate even at extremely low Kondo temperatures 
$\sim 10^{-12}$ of the band width, and our results are consistent with $x$ vanishing at zero Kondo coupling 
($U\to\infty$).

The exponential decay of correlation matrix eigenvalues (natural orbital occupation numbers) to
full occupancy or vacancy therefore is not a special feature of the IRLM representation of
Kondo physics. Thus, also for the SIAM, the single-particle Hilbert space can be partitioned into
an $M$-dimensional correlated sector, and a remainder that is uncorrelated. An ansatz that
straightforwardly generalizes (\ref{FewBodyWF}) can be constructed. Its accuracy 
is controlled by $M$, and any desired accuracy can be obtained with an $M$ that remains
finite in the thermodynamic limit. Due to the fact that there is an extra degeneracy 
in the correlation matrix spectrum of the SIAM and also because for given $T_K$, the decay
rate $x$ is roughly twice smaller for SIAM fermions than for IRLM fermions, a larger $M$ will 
be required for the same accuracy at a given $T_K$ than in the IRLM. As a practical matter, this limits the
range of Kondo couplings for which few-body approximations to the ground state of the SIAM
can be found numerically, but in principle, the SIAM ground state is effectively a few-body correlated state
in terms of SIAM fermions, in the same way that the IRLM ground state is few-body in nature,
provided $T_K$ is finite.

\section{Conclusions}
\label{s5}
We have calculated the correlation matrix of the IRLM and the SIAM, two quantum impurity 
models 
that are equivalent to the single channel Kondo Hamiltonian.
Several recent studies have 
noted that the eigenvalues of the correlation matrix of quantum impurity models often
decay exponentially towards full occupation or vacancy\cite{He2014,Lu2014,Zheng2020}, and our results 
confirm this observation, provided that the ground state is not quantum critical. We have 
however identified results in the literature about the Kondo model, namely
that the exponential decay rate of correlation matrix eigenvalues become
large close to the weak coupling critical point, that are finite size artefacts.
We demonstrated that, in fact, the decay rate tends to zero as the critical
point is approached for a macroscopically large electronic bath. Finite
size systems that are smaller than the Kondo length prevent the full development of correlations.
We have also investigated the spatial structure of the most correlated
natural orbitals as the critical point is approached and detected clear fingerprints
of the Kondo screening cloud.

Our main result presents a general method for determining the effective number 
of correlated particles (on top of an uncorrelated Fermi sea). 
 This involves using the natural orbital single-particle basis
to identify correlated and uncorrelated sectors of Fock space. Owing to the
exponential decay of the correlation matrix spectrum to full occupancy or vacancy,
the correlated sector can, to a good approximation, be chosen to contain a finite number 
of particles $M/2$, whereas the uncorrelated sector contains an infinite number
of particles within a single Slater determinant in the thermodynamic limit.
The full ground state can be reconstructed approximately
by solving an effective few-body problem for the particles in the correlated sector.
If the reconstructed state has an energy expectation value that differs from the
true ground state by an amount that is significantly less than the Kondo temperature,
then the reconstructed state is a faithful approximation of the true ground state. 
By comparing the energy of this reconstructed state to the true ground state energy,
as a function of $M$, we can thus determine the effective number of correlated particles.
Whereas the number of correlated particles diverge at the weak coupling fixed point 
($T_{\rm K}\to0$),
for realistic Kondo temperatures of $\sim 10^{-3}$ of the Fermi energy, the ground state
only hosts around seven correlated particles in the IRLM representation, and a larger
but still finite number in the SIAM. The different models host different numbers of
correlated particles at the same Kondo temperature because their microscopic degrees
of freedom are nontrivially related.
We have investigated how this picture is affected when correlations are
frustrated, either by finite size effects, or by symmetry-breaking fields.
We showed that, as expected, physical cutoffs acting near the Kondo scale
are accompanied by a sharp reduction in the number of correlated particles. 
However, we anticipate that models tuned to criticality, 
such as the two-channel and two-impurity Kondo models, remain truly many-body in any 
single-particle basis.
Our results open many interesting avenues for research, such as generalizations
to other quantum impurity problems or even to disordered lattice models.
It would also be interesting to investigate whether this few-body picture is robust 
for excited or unitarily time-evolved states, a notoriously challenging problem
for strongly interacting fermions. 

\acknowledgments We thank M.-B. Lepetit for discussions, the National Research Foundation of 
South Africa (Grant No. 90657), and the CNRS PICS contract FERMICATS for support.

\appendix

\section{Convergence of the spectrum of $\hat{Q}$ to the thermodynamic limit.}
\label{appa}

\begin{figure*}[htb]
\includegraphics[width=1.5 \columnwidth]{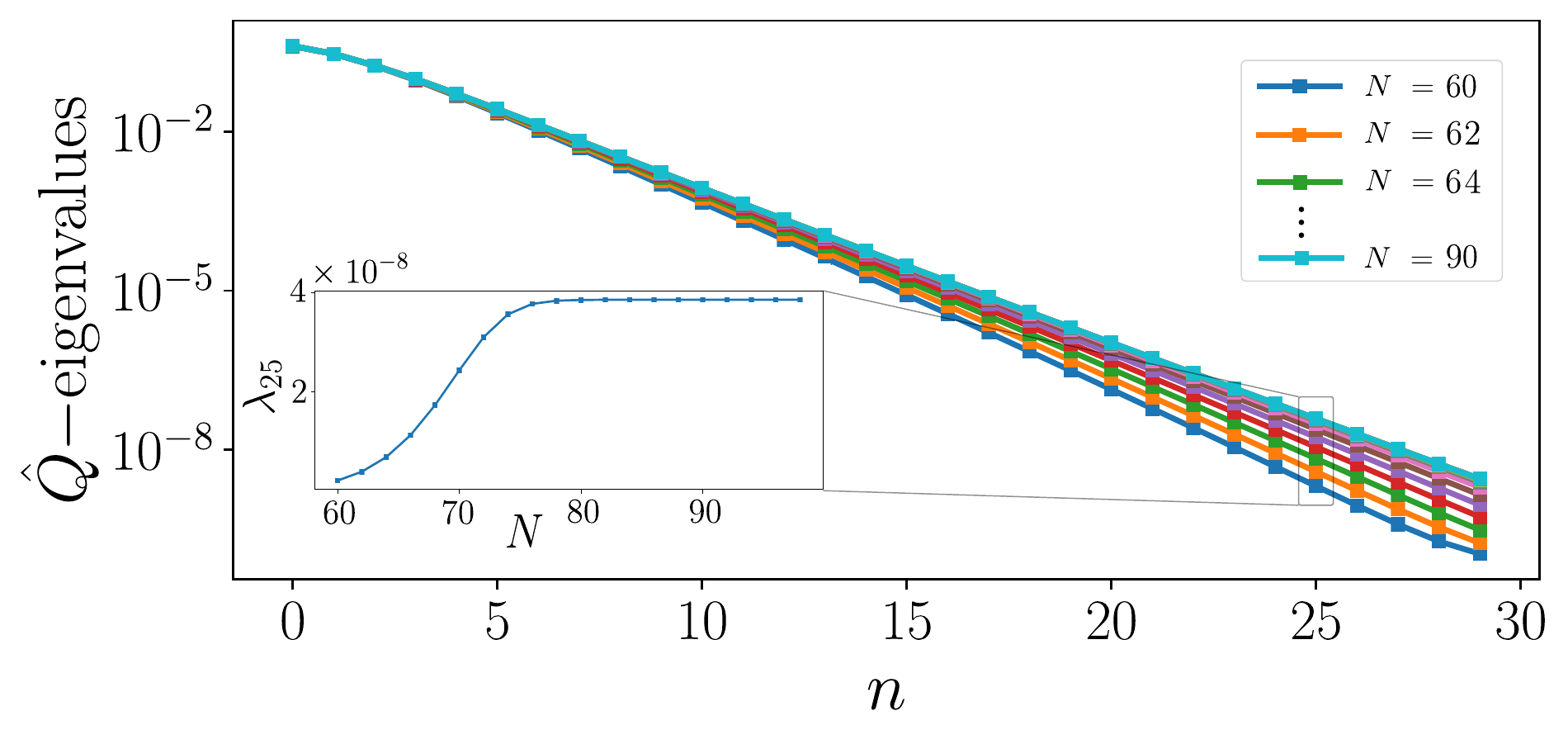}
\includegraphics[width=0.99\columnwidth]{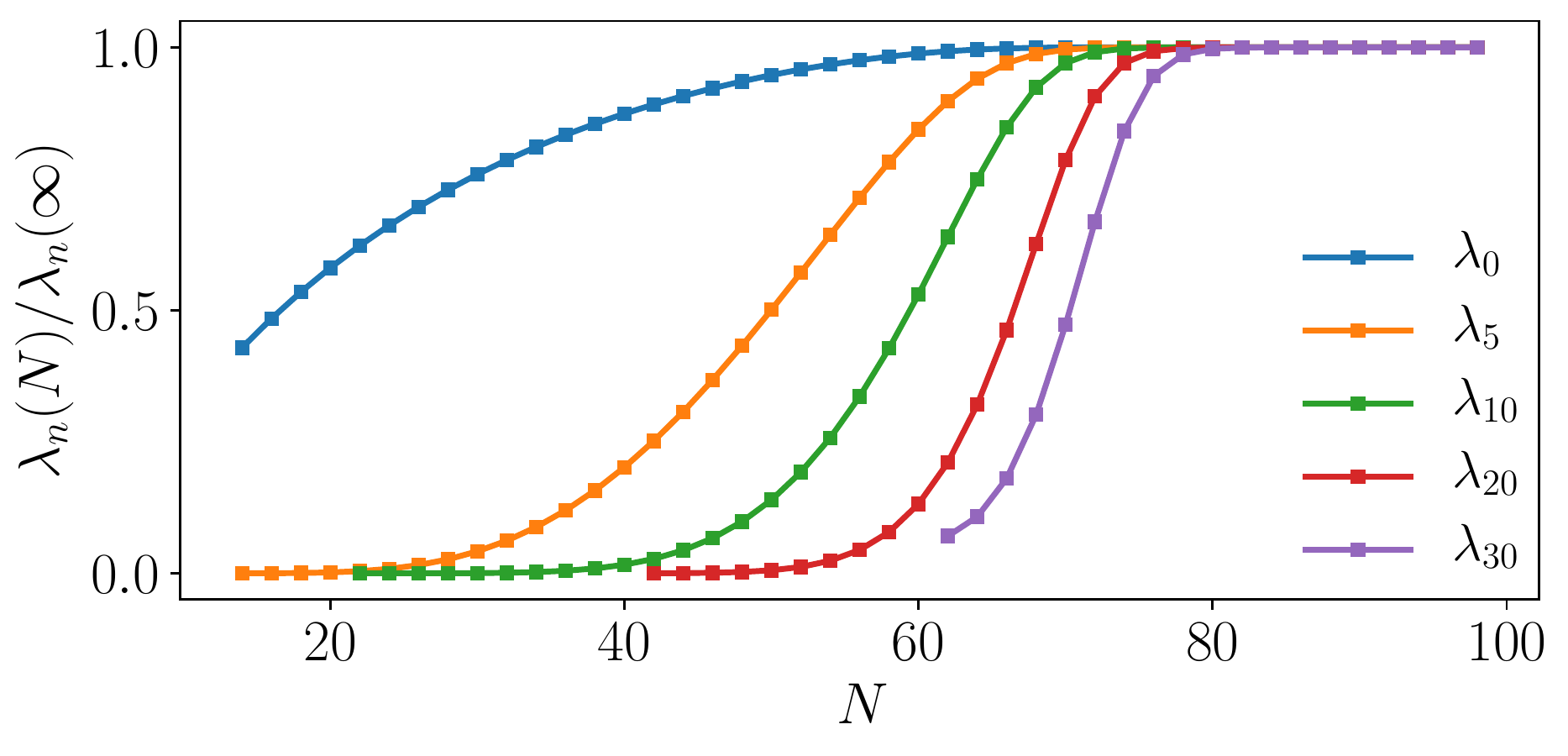}
\includegraphics[width=0.99\columnwidth]{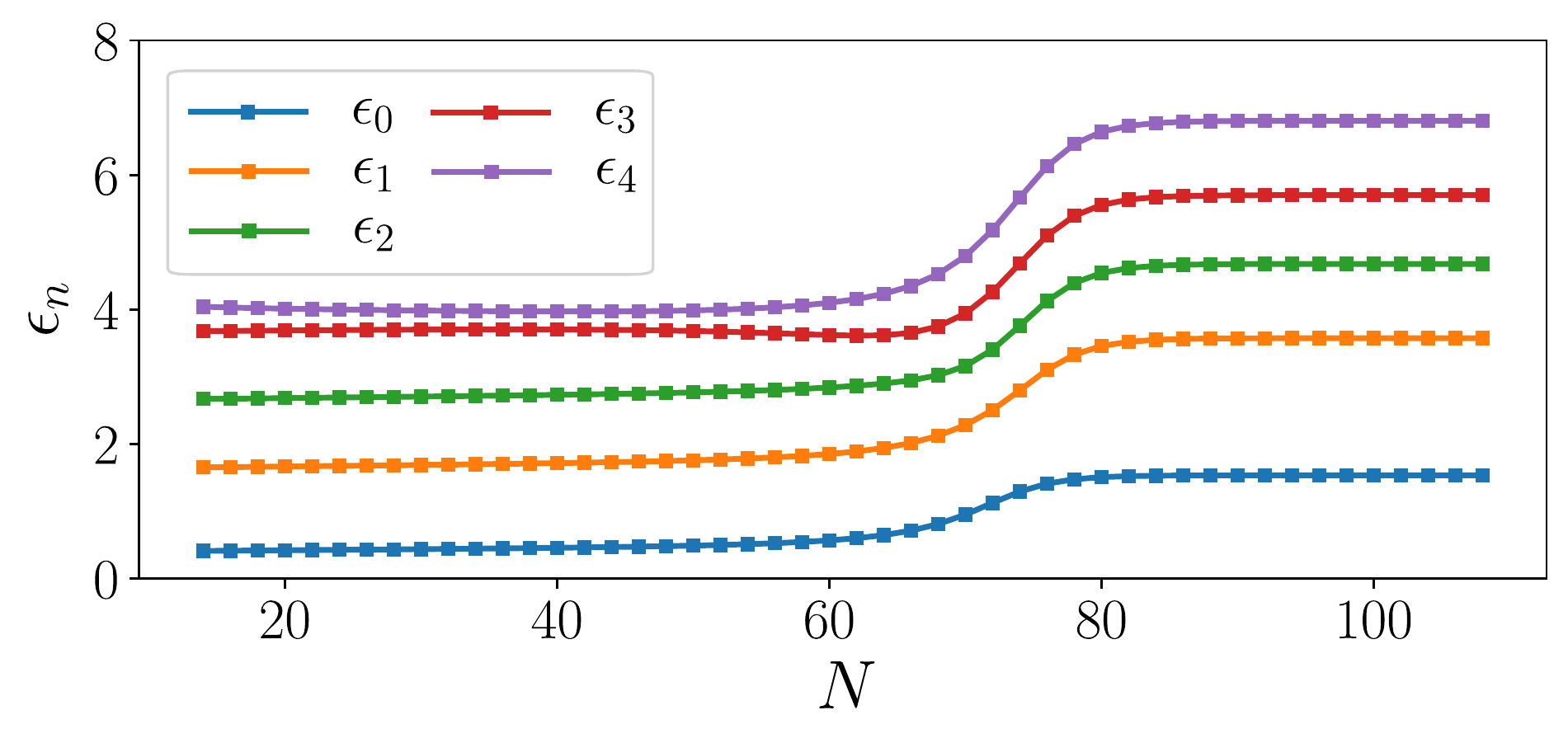}
\caption{Top: $\hat{Q}$-matrix spectrum for $U=-1.2D$ near the quantum critical point
$U_c=-1.3D$ for various chain lengths $N$ ($\Lambda=2.25$ here). Bottom left: 
same data, plotted as a function of $N$, showing convergence for $N>75$.
Bottom right: corresponding flow of the five lowest eigenvalues of the 
rescaled Hamiltonian. Both the $\hat{Q}$-eigenvalues $\lambda_n(N)$ and the H-eigenvalues
$\epsilon_n(N)$ show a crossover to the Kondo fixed point at a scale $N\gtrsim75$, 
corresponding to $T_K/D\simeq2.25^{-75/2}\simeq10^{-13}$.}
\label{ConvN}
\end{figure*}

The thermodynamic limit of the Wilson chain used in NRG is obtained mathematically by sending the chain length $N$ to infinity
and subsequently sending $\Lambda$ to $1$. For the IRLM, this limit describes a one-dimensional conduction band with a constant
density of states coupled via hybridization and short range Coulomb interactions to a resonant level.
In practice, numerical calculations are performed at finite $N$ and $\Lambda>1$, and also introduce a further regularization
parameter, $N_{\rm kept}$, the maximum dimension to which fixed particle number sectors of Hilbert space
are truncated in each renormalization step. In this section we demonstrate that our numerical results are converged
to the thermodynamic limit with respect to these three regularization parameters.
We first show in Fig.~\ref{ConvN} (top panel) that the $\hat{Q}$-matrix spectrum is indeed well converged for
suffienciently long chains. Here, we consider an interaction
value $U=-1.2D$ very close to the quantum critical point $U_c=-1.3D$, leading
to an exponentially small Kondo temperature of order $T_K/D\simeq\Lambda^{-75/2}\simeq 10^{-13}$,
which is estimated from the crossover at $N\simeq75$ seen in the 
flow of the lowest eigenvalues of the rescaled Hamiltonian
(bottom right panel in Fig.~\ref{ConvN}), using the value $\Lambda=2.25$ for
this NRG computation.
The same crossover scale is seen for all eigenvalues $\lambda_n$ (see bottom
left panel in Fig.~\ref{ConvN}), which are well saturated to their $N=\infty$ limit
for $N>75$. Note however, that eigenvalues $\lambda_n$ with $n>N/2$ are not
defined since the chain is too short to harbor those modes. Longer chains are
thus required to obtain such small eigenvalues.

\begin{figure*}[htb]
\includegraphics[width=0.65\columnwidth]{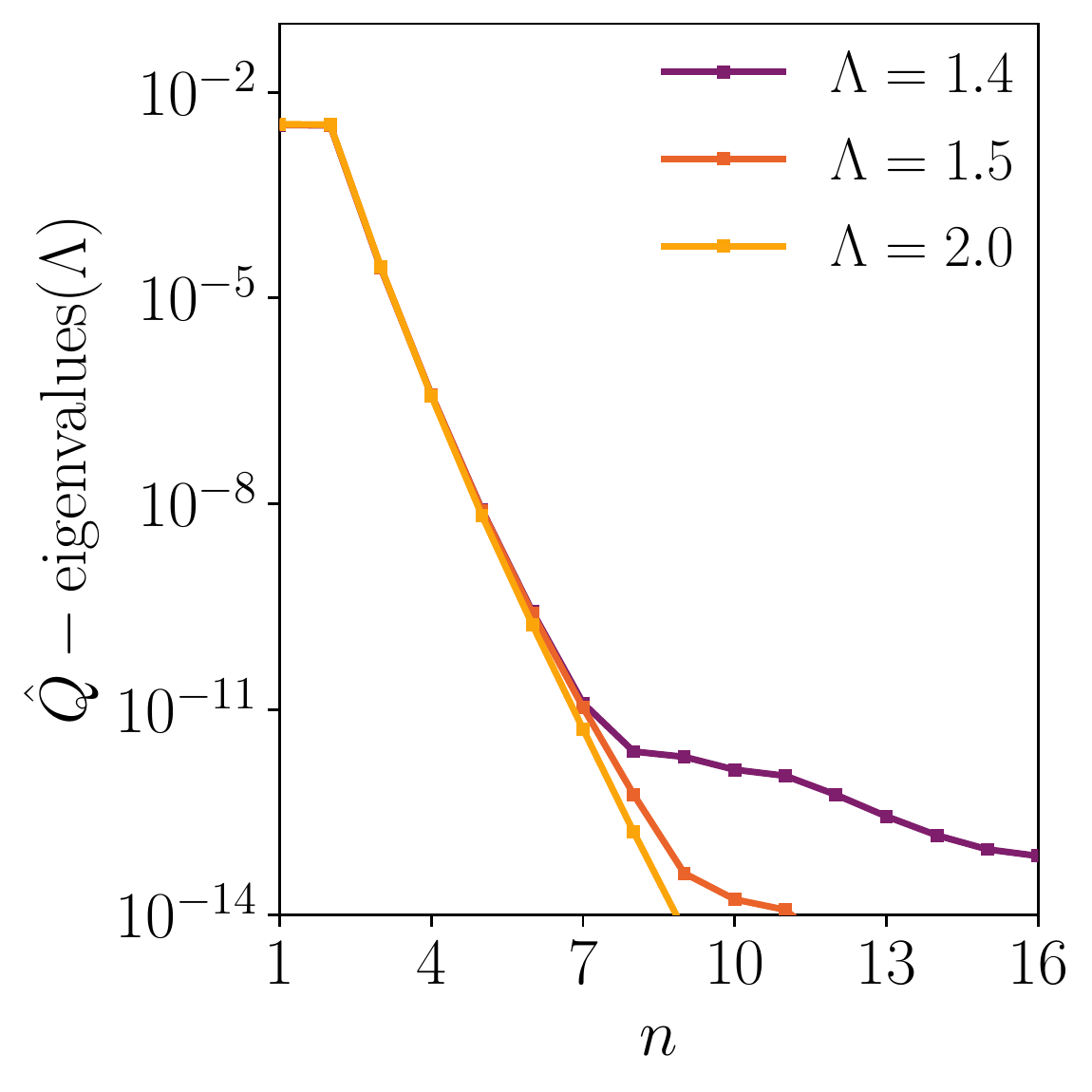}
\includegraphics[width=0.65\columnwidth]{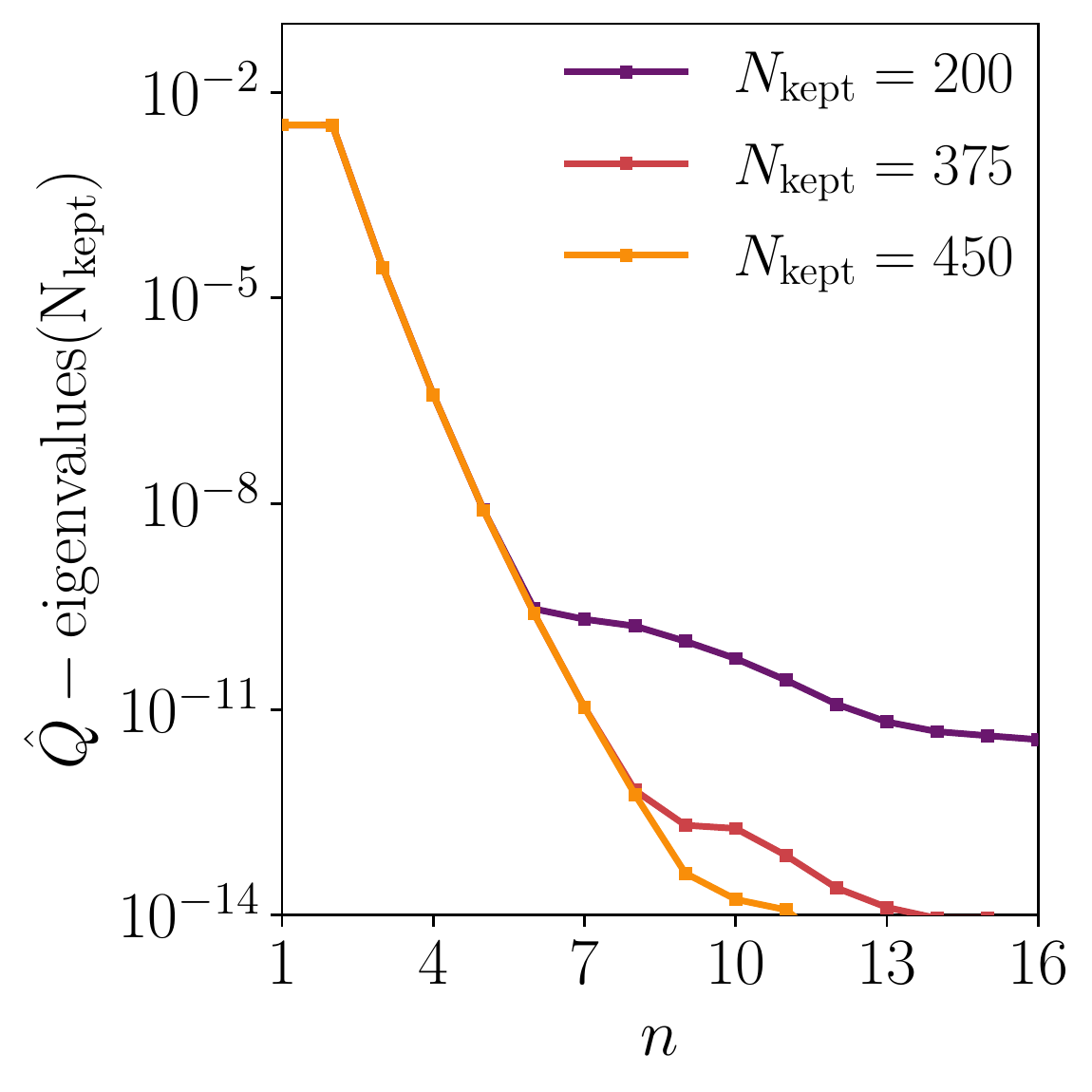}
\includegraphics[width=0.65\columnwidth]{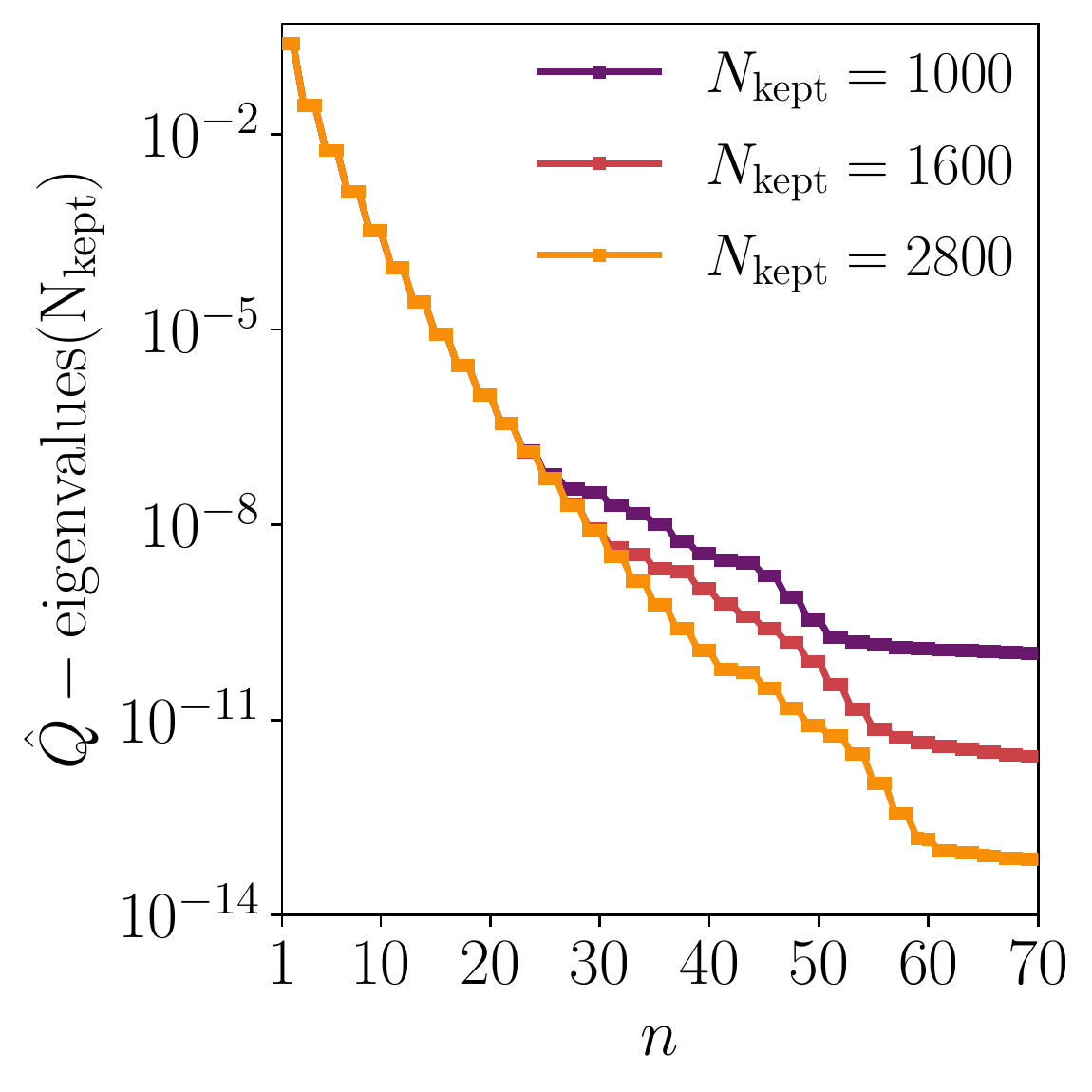}
\caption{Left: IRLM $\hat{Q}$-spectrum for $U=0.5$ and $N_\mr{kept}=450$, for
$\Lambda=1.4,1.5,2.0$, showing robustness of the exponential decay in
the thermodynamic limit. Center: IRLM $\hat{Q}$-spectrum as a function of $N_\mr{kept}$ 
for $\Lambda=1.5$, displaying typical truncation errors of the NRG, and their disappearance when
increasing the number of kept states. Right: SIAM $\hat{Q}$-spectrum versus
$N_\mr{kept}$, for $\Gamma = 0.01D$, $U=0.2D=20\Gamma$ and $\Lambda = 2.0$.}
\label{sf_qev}
\end{figure*}

We then investigate the issue of the convergence to the continuum limit
$\Lambda\to1$. In the left panel of Fig.~\ref{sf_qev}, we plot the $\hat{Q}$-spectrum
for $U=0.5$, with $N_\mr{kept}=110$ many-body states per block, for various
$\Lambda$ values. The first 5 eigenvalues are clearly independent of $\Lambda$,
showing that the exponential falloff is robust in the thermodynamic limit.
For the smaller eigenvalues $\lambda_n$ with $n\geq6$, some departure of the
exponential decay is seen for $\Lambda=1.5$ and $\Lambda=1.4$. We show in the
middle panel of Fig.~\ref{sf_qev} that this artefact is purely an effect of the
truncation error on the exponentially small magnitude of the eigenvalues, that
disappears progressively when increasing $N_\mr{kept}$.
Thus, in practice, calculations with $\Lambda=2$ and $N_\mr{kept}\simeq100$
provide good convergence for spinless models, emphasizing that it is not useful 
to consider $\lambda_n$ eigenvalues below the machine precision $10^{-16}$.
We found similar results for other values of the interaction $U$.

In Sec. \ref{ssiam} we presented results for the spectrum of the correlation
matrix of the single impurity Anderson model.
Given the larger single particle Hilbert space, it is important to make sure that these results are
converged with respect to NRG truncation. In the right panel of Fig.\,\ref{sf_qev}, we show that this 
is the case: The evolution of the spectrum of the numerically computed correlation matrix for the 
SIAM as a function of $N_{\rm kept}$ 
closely mirrors that of the IRLM (compare to the middle panel of Fig.\,\ref{sf_qev}.) 
We clearly see that the truncation error is pushed closer and closer to full occupancy or vacancy
as $N_{\rm kept}$ is increased, and that results are consistent with a spectrum that decays
exponentially to full occupancy or vacancy.

\section{Few-body diagonalization in the $\hat{Q}$-eigenbasis}
\label{appb}

We present here some technical details on how to perform an exact
diagonalization of an arbitrary non-local Hamiltonian, using the eigenvectors 
of the $\hat{Q}$-matrix as an optimized set of $M$ correlated orbitals $q^\dagger_n$, with
$n=-M/2,\ldots-1,1,\ldots,M/2$. The orbitals with $-N/2\leq
n<-M/2$ are fully occupied, while orbitals with $M/2<n\leq N/2$ are totally
empty. As discussed in the main text, we write the full wave function as follows:
\begin{eqnarray}
|\Psi_\mr{few}\rangle &=&
{\sum_{\{N_n=0,1\}}}
\Psi(N_{-\frac{M}{2}},\ldots, N_{\frac{M}{2}})\!\!\!\!
\prod_{n=-\frac{M}{2}}^{\frac{M}{2}}\!\!
[q^{\dagger}_n]^{N_n}
|\Psi_0\rangle,
\label{sFewBodyWF}\nonumber\\
\left|\Psi_0\right>&=&\prod_{m=-\frac{N}{2}}^{-\frac{M}{2}-1}
q^{\dagger}_{m}\left|0\right>,
\end{eqnarray}
with $N_n=0,1$ the occupancy of correlated orbital $q^\dagger_n$, the summation
restricted to occupations such that $\sum_{n=-M/2}^{M/2}N_n=M/2$ at half-filling, 
and $\Psi(N_{-\frac{M}{2}},\ldots, N_{\frac{M}{2}})$ the complete wave function
in the correlated subspace.

We then split the full Hamiltonian between the correlated and uncorrelated
sectors as ${\cal{H}} = {\cal{H}}_\mr{corr} + {\cal{H}}_\mr{mix} +
{\cal{H}}_\mr{uncorr}$.
We label the correlated orbitals with roman indices, such as $q^\dagger_n$ with
$n=-M/2,\ldots,-1,1,\ldots,M/2$, and
uncorrelated orbitals with greek indices, such as $q^\dagger_\alpha$ with
$\alpha=-N/2,\ldots,-M/2-1,M/2+1,\ldots N/2$. The various terms read:

\begin{equation}
\begin{split}
{\cal{H}}_\mr{corr} &= \sum\limits_{n,m}t_{nm}q^{\dagger}_{n} q^{}_{m} + \sum\limits_{n,m,p,q}U_{nmpq}\;
q^{\dagger}_{n}q^{\dagger}_{m}q^{}_{p}q^{ }_{q}\\
{\cal{H}}_\mr{mix} &= 
\sum\limits_{n,m}\sum\limits_{\alpha,\beta} \Big( 
U_{n,\alpha,m,\beta} \; q^{\dagger}_{n}q^{\dagger}_{\alpha} q^{}_{m}q^{}_{\beta}\\
&~~~~~~~~~~~~+ U_{n,\alpha,\beta,m} \; q^{\dagger}_{n}q^{\dagger}_{\alpha} q^{}_{\beta}q^{}_{m}\\ 
&~~~~~~~~~~~~+ U_{\alpha,n,m,\beta} \; q^{\dagger}_{\alpha}q^{\dagger}_{n} q^{}_{m}q^{}_{\beta}\\
&~~~~~~~~~~~~+ U_{\alpha,n,\beta,m} \; q^{\dagger}_{\alpha}q^{\dagger}_{n} q^{}_{\beta}q^{}_{m} \Big)
+ {\cal{H}}_\mr{odd}\\
{\cal{H}}_\mr{uncorr} &=
\sum\limits_{\alpha,\beta}t_{\alpha\beta}q^{\dagger}_{\alpha} q^{}_{\beta} 
+ \sum\limits_{\alpha,\beta,\gamma,\delta}U_{\alpha\beta\gamma\delta}\;
q^{\dagger}_{\alpha}q^{\dagger}_{\beta}q^{}_{\gamma}q^{ }_{\delta},
\end{split}
\end{equation}
where ${\cal{H}}_\mr{odd}$ contains odd terms in the uncorrelated orbitals, such
as the hopping term $q^\dagger_\alpha q^{}_n$ mixing both sectors, or
interaction terms of the form $q^\dagger_\alpha q^\dagger_nq^{}_m q^{}_p$. The terms in ${\cal{H}}_\mr{odd}$
vansih once we project the Hamiltonian in the family of
states of the form~(\ref{sFewBodyWF}).
The resulting effective few-body Hamiltonian reads:
\begin{eqnarray}
\nonumber
{\cal{H}}_\mr{few} &=& \sum\limits_{n,m}t_{nm}q^{\dagger}_{n} q^{}_{m} + \sum\limits_{n,m,p,q}U_{nmpq}\;
q^{\dagger}_{n}q^{\dagger}_{m}q^{}_{p}q^{ }_{q}\\
&&
+ \sum\limits_{n,m}\sum\limits_{\alpha} q^{\dagger}_{n}q^{}_{m} n_\alpha 
\Big(- U_{n,\alpha,m,\alpha} + U_{n,\alpha,\alpha,m}\nonumber\\
&&~~~~~~~~~~~~~~~~~~~~~~~~ + U_{\alpha,n,m,\alpha} -
U_{\alpha,n,\alpha,m}\Big)\nonumber\\
&& + \sum\limits_{\alpha}t_{\alpha\alpha}n_\alpha
+ \sum\limits_{\alpha\neq\beta}(U_{\alpha\beta\beta\alpha}-U_{\alpha\beta\alpha\beta})n_\alpha
n_\beta,\nonumber\\
\end{eqnarray}
where $n_\alpha$ is the occupancy of the uncorrelated orbitals in the 
wavefunction~(\ref{sFewBodyWF}), namely $n_\alpha=1$ for $\alpha<-M/2$ 
and $n_\alpha=0$ for $\alpha>M/2$. We note that projecting ${\cal{H}}_\mr{mix}$
generates a renormalization of the hopping term $q^\dagger_n q^{}_m$ within the 
correlated sector (last term under parenthesis in the first line of the equation above).
The projection of ${\cal{H}}_\mr{uncorr}$ provides only a constant contribution to
the Hamiltonian (second line of the equation above). The ground state energy of
the initial many-body Hamiltonian ${\cal{H}}$ is obtained by exact
diagonalization of ${\cal{H}}_\mr{few}$ in the few-body correlated sector.

\section{Extraction of the Kondo temperature}
\label{appc}
\begin{figure}[htb]
\includegraphics[width=0.99\columnwidth]{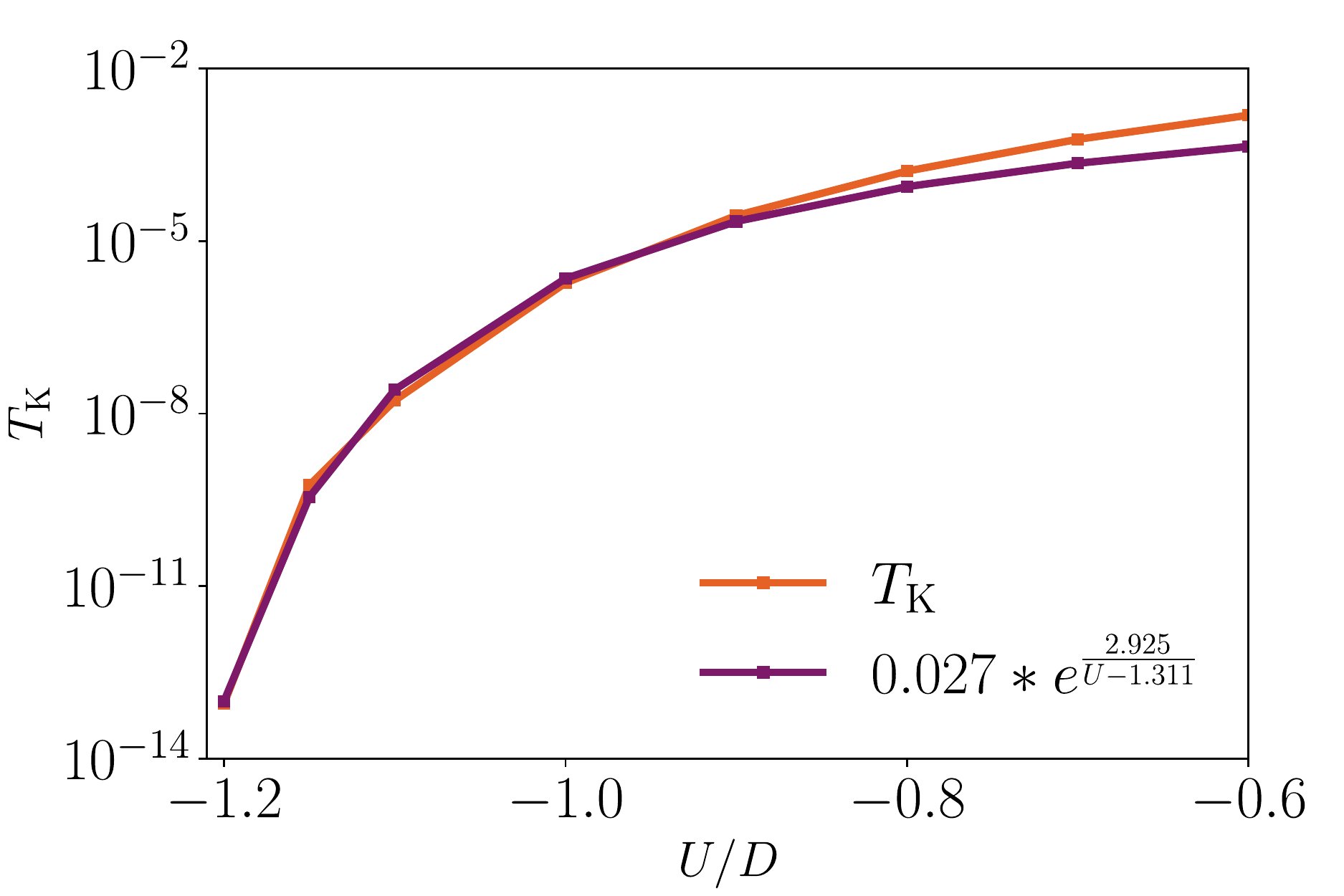}
\caption{Kondo temperature of the IRLM as a function of interaction $U$
computed by NRG with Wilson discretization parameter $\Lambda=2$ and chains
of length up to $N=110$.}
\label{f_TK}
\end{figure}

For the IRLM, we take the following definition for the Kondo temperature:
\begin{eqnarray}
T_K&=&\frac{1}{4\chi} \\
\chi&=&\lim_{\epsilon_d\to0} \frac{\mr{d}}{\mr{d}\epsilon_d}
\big<d^\dagger d\big>,
\end{eqnarray}
upon adding to the IRLM Hamiltonian a local potential on the $d$-level,
namely a term $\epsilon_d d^\dagger d$. This is equivalent to
the standard definition \cite{Weichselbaum2014} for the Kondo model in terms of the magnetic susceptibility of the
impurity spin. The resulting Kondo $T_K$ temperature
as a function of interaction $U$ is given in Fig.~\ref{f_TK}. A fit of the
essential singularity at the critical point allows to determine the critical
value $U_c\simeq-1.3$. We also recover our previous estimate $T_K/D\simeq10^{-13}$
at $U=-1.2$ shown previously in Fig.~\ref{ConvN}.


\begin{thebibliography}{31}%
\makeatletter
\providecommand \@ifxundefined [1]{%
 \@ifx{#1\undefined}
}%
\providecommand \@ifnum [1]{%
 \ifnum #1\expandafter \@firstoftwo
 \else \expandafter \@secondoftwo
 \fi
}%
\providecommand \@ifx [1]{%
 \ifx #1\expandafter \@firstoftwo
 \else \expandafter \@secondoftwo
 \fi
}%
\providecommand \natexlab [1]{#1}%
\providecommand \enquote  [1]{``#1''}%
\providecommand \bibnamefont  [1]{#1}%
\providecommand \bibfnamefont [1]{#1}%
\providecommand \citenamefont [1]{#1}%
\providecommand \href@noop [0]{\@secondoftwo}%
\providecommand \href [0]{\begingroup \@sanitize@url \@href}%
\providecommand \@href[1]{\@@startlink{#1}\@@href}%
\providecommand \@@href[1]{\endgroup#1\@@endlink}%
\providecommand \@sanitize@url [0]{\catcode `\\12\catcode `\$12\catcode
  `\&12\catcode `\#12\catcode `\^12\catcode `\_12\catcode `\%12\relax}%
\providecommand \@@startlink[1]{}%
\providecommand \@@endlink[0]{}%
\providecommand \url  [0]{\begingroup\@sanitize@url \@url }%
\providecommand \@url [1]{\endgroup\@href {#1}{\urlprefix }}%
\providecommand \urlprefix  [0]{URL }%
\providecommand \Eprint [0]{\href }%
\providecommand \doibase [0]{https://doi.org/}%
\providecommand \selectlanguage [0]{\@gobble}%
\providecommand \bibinfo  [0]{\@secondoftwo}%
\providecommand \bibfield  [0]{\@secondoftwo}%
\providecommand \translation [1]{[#1]}%
\providecommand \BibitemOpen [0]{}%
\providecommand \bibitemStop [0]{}%
\providecommand \bibitemNoStop [0]{.\EOS\space}%
\providecommand \EOS [0]{\spacefactor3000\relax}%
\providecommand \BibitemShut  [1]{\csname bibitem#1\endcsname}%
\let\auto@bib@innerbib\@empty
%</preamble>
\bibitem [{\citenamefont {Krishna-murthy}\ \emph {et~al.}(1980)\citenamefont
  {Krishna-murthy}, \citenamefont {Wilkins},\ and\ \citenamefont
  {Wilson}}]{Wilson_1980}%
  \BibitemOpen
  \bibfield  {author} {\bibinfo {author} {\bibfnamefont {H.~R.}\ \bibnamefont
  {Krishna-murthy}}, \bibinfo {author} {\bibfnamefont {J.~W.}\ \bibnamefont
  {Wilkins}},\ and\ \bibinfo {author} {\bibfnamefont {K.~G.}\ \bibnamefont
  {Wilson}},\ }\bibfield  {title} {\bibinfo {title} {Renormalization-group
  approach to the anderson model of dilute magnetic alloys. i. static
  properties for the symmetric case},\ }\href
  {https://doi.org/10.1103/PhysRevB.21.1003} {\bibfield  {journal} {\bibinfo
  {journal} {Phys. Rev. B}\ }\textbf {\bibinfo {volume} {21}},\ \bibinfo
  {pages} {1003} (\bibinfo {year} {1980})}\BibitemShut {NoStop}%
\bibitem [{\citenamefont {White}(1992)}]{White_DMRG}%
  \BibitemOpen
  \bibfield  {author} {\bibinfo {author} {\bibfnamefont {S.~R.}\ \bibnamefont
  {White}},\ }\bibfield  {title} {\bibinfo {title} {Density matrix formulation
  for quantum renormalization groups},\ }\href
  {https://doi.org/10.1103/PhysRevLett.69.2863} {\bibfield  {journal} {\bibinfo
   {journal} {Phys. Rev. Lett.}\ }\textbf {\bibinfo {volume} {69}},\ \bibinfo
  {pages} {2863} (\bibinfo {year} {1992})}\BibitemShut {NoStop}%
\bibitem [{\citenamefont {Gull}\ \emph {et~al.}(2011)\citenamefont {Gull},
  \citenamefont {Millis}, \citenamefont {Lichtenstein}, \citenamefont
  {Rubtsov}, \citenamefont {Troyer},\ and\ \citenamefont {Werner}}]{Gull}%
  \BibitemOpen
  \bibfield  {author} {\bibinfo {author} {\bibfnamefont {E.}~\bibnamefont
  {Gull}}, \bibinfo {author} {\bibfnamefont {A.~J.}\ \bibnamefont {Millis}},
  \bibinfo {author} {\bibfnamefont {A.~I.}\ \bibnamefont {Lichtenstein}},
  \bibinfo {author} {\bibfnamefont {A.~N.}\ \bibnamefont {Rubtsov}}, \bibinfo
  {author} {\bibfnamefont {M.}~\bibnamefont {Troyer}},\ and\ \bibinfo {author}
  {\bibfnamefont {P.}~\bibnamefont {Werner}},\ }\bibfield  {title} {\bibinfo
  {title} {Continuous-time monte carlo methods for quantum impurity models},\
  }\href {https://doi.org/10.1103/RevModPhys.83.349} {\bibfield  {journal}
  {\bibinfo  {journal} {Rev. Mod. Phys.}\ }\textbf {\bibinfo {volume} {83}},\
  \bibinfo {pages} {349} (\bibinfo {year} {2011})}\BibitemShut {NoStop}%
\bibitem [{\citenamefont {Georges}\ \emph {et~al.}(1996)\citenamefont
  {Georges}, \citenamefont {Kotliar}, \citenamefont {Krauth},\ and\
  \citenamefont {Rozenberg}}]{Georges1996}%
  \BibitemOpen
  \bibfield  {author} {\bibinfo {author} {\bibfnamefont {A.}~\bibnamefont
  {Georges}}, \bibinfo {author} {\bibfnamefont {G.}~\bibnamefont {Kotliar}},
  \bibinfo {author} {\bibfnamefont {W.}~\bibnamefont {Krauth}},\ and\ \bibinfo
  {author} {\bibfnamefont {M.~J.}\ \bibnamefont {Rozenberg}},\ }\bibfield
  {title} {\bibinfo {title} {Dynamical mean-field theory of strongly correlated
  fermion systems and the limit of infinite dimensions},\ }\href
  {https://doi.org/10.1103/RevModPhys.68.13} {\bibfield  {journal} {\bibinfo
  {journal} {Rev. Mod. Phys.}\ }\textbf {\bibinfo {volume} {68}},\ \bibinfo
  {pages} {13} (\bibinfo {year} {1996})}\BibitemShut {NoStop}%
\bibitem [{\citenamefont {Eisert}\ \emph {et~al.}(2010)\citenamefont {Eisert},
  \citenamefont {Cramer},\ and\ \citenamefont {Plenio}}]{Entanglement_Review}%
  \BibitemOpen
  \bibfield  {author} {\bibinfo {author} {\bibfnamefont {J.}~\bibnamefont
  {Eisert}}, \bibinfo {author} {\bibfnamefont {M.}~\bibnamefont {Cramer}},\
  and\ \bibinfo {author} {\bibfnamefont {M.~B.}\ \bibnamefont {Plenio}},\
  }\bibfield  {title} {\bibinfo {title} {Colloquium: Area laws for the
  entanglement entropy},\ }\href {https://doi.org/10.1103/RevModPhys.82.277}
  {\bibfield  {journal} {\bibinfo  {journal} {Rev. Mod. Phys.}\ }\textbf
  {\bibinfo {volume} {82}},\ \bibinfo {pages} {277} (\bibinfo {year}
  {2010})}\BibitemShut {NoStop}%
\bibitem [{\citenamefont {\"Ostlund}\ and\ \citenamefont
  {Rommer}(1995)}]{OstlundRommer}%
  \BibitemOpen
  \bibfield  {author} {\bibinfo {author} {\bibfnamefont {S.}~\bibnamefont
  {\"Ostlund}}\ and\ \bibinfo {author} {\bibfnamefont {S.}~\bibnamefont
  {Rommer}},\ }\bibfield  {title} {\bibinfo {title} {Thermodynamic limit of
  density matrix renormalization},\ }\href
  {https://doi.org/10.1103/PhysRevLett.75.3537} {\bibfield  {journal} {\bibinfo
   {journal} {Phys. Rev. Lett.}\ }\textbf {\bibinfo {volume} {75}},\ \bibinfo
  {pages} {3537} (\bibinfo {year} {1995})}\BibitemShut {NoStop}%
\bibitem [{\citenamefont {{Dukelsky, J.}}\ \emph {et~al.}(1998)\citenamefont
  {{Dukelsky, J.}}, \citenamefont {{Mart\'{\i}n-Delgado, M. A.}}, \citenamefont
  {{Nishino, T.}},\ and\ \citenamefont {{Sierra, G.}}}]{Dukelsky1998}%
  \BibitemOpen
  \bibfield  {author} {\bibinfo {author} {\bibnamefont {{Dukelsky, J.}}},
  \bibinfo {author} {\bibnamefont {{Mart\'{\i}n-Delgado, M. A.}}}, \bibinfo
  {author} {\bibnamefont {{Nishino, T.}}},\ and\ \bibinfo {author}
  {\bibnamefont {{Sierra, G.}}},\ }\bibfield  {title} {\bibinfo {title}
  {Equivalence of the variational matrix product method and the density matrix
  renormalization group applied to spin chains},\ }\href
  {https://doi.org/10.1209/epl/i1998-00381-x} {\bibfield  {journal} {\bibinfo
  {journal} {Europhys. Lett.}\ }\textbf {\bibinfo {volume} {43}},\ \bibinfo
  {pages} {457} (\bibinfo {year} {1998})}\BibitemShut {NoStop}%
\bibitem [{\citenamefont {Vidal}(2003)}]{Vidal2003}%
  \BibitemOpen
  \bibfield  {author} {\bibinfo {author} {\bibfnamefont {G.}~\bibnamefont
  {Vidal}},\ }\bibfield  {title} {\bibinfo {title} {Efficient classical
  simulation of slightly entangled quantum computations},\ }\href
  {https://doi.org/10.1103/PhysRevLett.91.147902} {\bibfield  {journal}
  {\bibinfo  {journal} {Phys. Rev. Lett.}\ }\textbf {\bibinfo {volume} {91}},\
  \bibinfo {pages} {147902} (\bibinfo {year} {2003})}\BibitemShut {NoStop}%
\bibitem [{\citenamefont {Schollw{\"o}ck}(2011)}]{Schollwock_Review}%
  \BibitemOpen
  \bibfield  {author} {\bibinfo {author} {\bibfnamefont {U.}~\bibnamefont
  {Schollw{\"o}ck}},\ }\bibfield  {title} {\bibinfo {title} {The density-matrix
  renormalization group in the age of matrix product states},\ }\href
  {https://doi.org/https://doi.org/10.1016/j.aop.2010.09.012} {\bibfield
  {journal} {\bibinfo  {journal} {Annals of Physics}\ }\textbf {\bibinfo
  {volume} {326}},\ \bibinfo {pages} {96 } (\bibinfo {year} {2011})},\ \bibinfo
  {note} {january 2011 Special Issue}\BibitemShut {NoStop}%
\bibitem [{\citenamefont {Abanin}\ \emph {et~al.}(2019)\citenamefont {Abanin},
  \citenamefont {Altman}, \citenamefont {Bloch},\ and\ \citenamefont
  {Serbyn}}]{MBL_Review}%
  \BibitemOpen
  \bibfield  {author} {\bibinfo {author} {\bibfnamefont {D.~A.}\ \bibnamefont
  {Abanin}}, \bibinfo {author} {\bibfnamefont {E.}~\bibnamefont {Altman}},
  \bibinfo {author} {\bibfnamefont {I.}~\bibnamefont {Bloch}},\ and\ \bibinfo
  {author} {\bibfnamefont {M.}~\bibnamefont {Serbyn}},\ }\bibfield  {title}
  {\bibinfo {title} {Colloquium: Many-body localization, thermalization, and
  entanglement},\ }\href {https://doi.org/10.1103/RevModPhys.91.021001}
  {\bibfield  {journal} {\bibinfo  {journal} {Rev. Mod. Phys.}\ }\textbf
  {\bibinfo {volume} {91}},\ \bibinfo {pages} {021001} (\bibinfo {year}
  {2019})}\BibitemShut {NoStop}%
\bibitem [{\citenamefont {Hewson}(1993)}]{Hewson1993}%
  \BibitemOpen
  \bibfield  {author} {\bibinfo {author} {\bibfnamefont {A.~C.}\ \bibnamefont
  {Hewson}},\ }\href@noop {} {\emph {\bibinfo {title} {The Kondo problem to
  heavy fermions}}}\ (\bibinfo  {publisher} {Cambridge University Press},\
  \bibinfo {address} {Cambridge New York},\ \bibinfo {year} {1993})\BibitemShut
  {NoStop}%
\bibitem [{\citenamefont {Bulla}\ \emph {et~al.}(2008)\citenamefont {Bulla},
  \citenamefont {Costi},\ and\ \citenamefont {Pruschke}}]{Bulla2008}%
  \BibitemOpen
  \bibfield  {author} {\bibinfo {author} {\bibfnamefont {R.}~\bibnamefont
  {Bulla}}, \bibinfo {author} {\bibfnamefont {T.~A.}\ \bibnamefont {Costi}},\
  and\ \bibinfo {author} {\bibfnamefont {T.}~\bibnamefont {Pruschke}},\
  }\bibfield  {title} {\bibinfo {title} {Numerical renormalization group method
  for quantum impurity systems},\ }\href
  {https://doi.org/10.1103/RevModPhys.80.395} {\bibfield  {journal} {\bibinfo
  {journal} {Rev. Mod. Phys.}\ }\textbf {\bibinfo {volume} {80}},\ \bibinfo
  {pages} {395} (\bibinfo {year} {2008})}\BibitemShut {NoStop}%
\bibitem [{\citenamefont {Barzykin}\ and\ \citenamefont
  {Affleck}(1996)}]{Affleck1996}%
  \BibitemOpen
  \bibfield  {author} {\bibinfo {author} {\bibfnamefont {V.}~\bibnamefont
  {Barzykin}}\ and\ \bibinfo {author} {\bibfnamefont {I.}~\bibnamefont
  {Affleck}},\ }\bibfield  {title} {\bibinfo {title} {The kondo screening
  cloud: What can we learn from perturbation theory?},\ }\href
  {https://doi.org/10.1103/PhysRevLett.76.4959} {\bibfield  {journal} {\bibinfo
   {journal} {Phys. Rev. Lett.}\ }\textbf {\bibinfo {volume} {76}},\ \bibinfo
  {pages} {4959} (\bibinfo {year} {1996})}\BibitemShut {NoStop}%
\bibitem [{\citenamefont {Park}\ \emph {et~al.}(2013)\citenamefont {Park},
  \citenamefont {Lee}, \citenamefont {Oreg},\ and\ \citenamefont
  {Sim}}]{Park2013}%
  \BibitemOpen
  \bibfield  {author} {\bibinfo {author} {\bibfnamefont {J.}~\bibnamefont
  {Park}}, \bibinfo {author} {\bibfnamefont {S.-S.~B.}\ \bibnamefont {Lee}},
  \bibinfo {author} {\bibfnamefont {Y.}~\bibnamefont {Oreg}},\ and\ \bibinfo
  {author} {\bibfnamefont {H.-S.}\ \bibnamefont {Sim}},\ }\bibfield  {title}
  {\bibinfo {title} {How to directly measure a kondo cloud's length},\ }\href
  {https://doi.org/10.1103/PhysRevLett.110.246603} {\bibfield  {journal}
  {\bibinfo  {journal} {Phys. Rev. Lett.}\ }\textbf {\bibinfo {volume} {110}},\
  \bibinfo {pages} {246603} (\bibinfo {year} {2013})}\BibitemShut {NoStop}%
\bibitem [{\citenamefont {Barcza}\ \emph {et~al.}(2020)\citenamefont {Barcza},
  \citenamefont {Bauerbach}, \citenamefont {Eickhoff}, \citenamefont {Anders},
  \citenamefont {Gebhard},\ and\ \citenamefont {Legeza}}]{Barcza2020}%
  \BibitemOpen
  \bibfield  {author} {\bibinfo {author} {\bibfnamefont {G.}~\bibnamefont
  {Barcza}}, \bibinfo {author} {\bibfnamefont {K.}~\bibnamefont {Bauerbach}},
  \bibinfo {author} {\bibfnamefont {F.}~\bibnamefont {Eickhoff}}, \bibinfo
  {author} {\bibfnamefont {F.~B.}\ \bibnamefont {Anders}}, \bibinfo {author}
  {\bibfnamefont {F.}~\bibnamefont {Gebhard}},\ and\ \bibinfo {author}
  {\bibfnamefont {O.}~\bibnamefont {Legeza}},\ }\bibfield  {title} {\bibinfo
  {title} {Symmetric single-impurity kondo model on a tight-binding chain:
  Comparison of analytical and numerical ground-state approaches},\ }\href
  {https://doi.org/10.1103/PhysRevB.101.075132} {\bibfield  {journal} {\bibinfo
   {journal} {Phys. Rev. B}\ }\textbf {\bibinfo {volume} {101}},\ \bibinfo
  {pages} {075132} (\bibinfo {year} {2020})}\BibitemShut {NoStop}%
\bibitem [{\citenamefont {V.~Borzenets}\ \emph {et~al.}(2020)\citenamefont
  {V.~Borzenets}, \citenamefont {Shim}, \citenamefont {Chen}, \citenamefont
  {Ludwig}, \citenamefont {Wieck}, \citenamefont {Tarucha}, \citenamefont
  {Sim},\ and\ \citenamefont {Yamamoto}}]{Borzenets2020}%
  \BibitemOpen
  \bibfield  {author} {\bibinfo {author} {\bibfnamefont {I.}~\bibnamefont
  {V.~Borzenets}}, \bibinfo {author} {\bibfnamefont {J.}~\bibnamefont {Shim}},
  \bibinfo {author} {\bibfnamefont {J.~C.~H.}\ \bibnamefont {Chen}}, \bibinfo
  {author} {\bibfnamefont {A.}~\bibnamefont {Ludwig}}, \bibinfo {author}
  {\bibfnamefont {A.~D.}\ \bibnamefont {Wieck}}, \bibinfo {author}
  {\bibfnamefont {S.}~\bibnamefont {Tarucha}}, \bibinfo {author} {\bibfnamefont
  {H.~S.}\ \bibnamefont {Sim}},\ and\ \bibinfo {author} {\bibfnamefont
  {M.}~\bibnamefont {Yamamoto}},\ }\bibfield  {title} {\bibinfo {title}
  {Observation of the kondo screening cloud},\ }\href
  {https://doi.org/10.1038/s41586-020-2058-6} {\bibfield  {journal} {\bibinfo
  {journal} {Nature}\ }\textbf {\bibinfo {volume} {579}},\ \bibinfo {pages}
  {210} (\bibinfo {year} {2020})}\BibitemShut {NoStop}%
\bibitem [{\citenamefont {Yang}\ and\ \citenamefont
  {Feiguin}(2017)}]{Yang2017}%
  \BibitemOpen
  \bibfield  {author} {\bibinfo {author} {\bibfnamefont {C.}~\bibnamefont
  {Yang}}\ and\ \bibinfo {author} {\bibfnamefont {A.~E.}\ \bibnamefont
  {Feiguin}},\ }\bibfield  {title} {\bibinfo {title} {Unveiling the internal
  entanglement structure of the kondo singlet},\ }\href
  {https://doi.org/10.1103/PhysRevB.95.115106} {\bibfield  {journal} {\bibinfo
  {journal} {Phys. Rev. B}\ }\textbf {\bibinfo {volume} {95}},\ \bibinfo
  {pages} {115106} (\bibinfo {year} {2017})}\BibitemShut {NoStop}%
\bibitem [{\citenamefont {Zheng}\ \emph {et~al.}(2020)\citenamefont {Zheng},
  \citenamefont {He},\ and\ \citenamefont {Lu}}]{Zheng2020}%
  \BibitemOpen
  \bibfield  {author} {\bibinfo {author} {\bibfnamefont {R.}~\bibnamefont
  {Zheng}}, \bibinfo {author} {\bibfnamefont {R.}~\bibnamefont {He}},\ and\
  \bibinfo {author} {\bibfnamefont {Z.}~\bibnamefont {Lu}},\ }\bibfield
  {title} {\bibinfo {title} {Natural orbitals renormalization group approach to
  a kondo singlet},\ }\href {https://doi.org/10.1007/s11433-019-1520-3}
  {\bibfield  {journal} {\bibinfo  {journal} {Science China Physics, Mechanics
  \& Astronomy}\ }\textbf {\bibinfo {volume} {63}},\ \bibinfo {pages} {297411}
  (\bibinfo {year} {2020})}\BibitemShut {NoStop}%
\bibitem [{\citenamefont {He}\ and\ \citenamefont {Lu}(2014)}]{He2014}%
  \BibitemOpen
  \bibfield  {author} {\bibinfo {author} {\bibfnamefont {R.-Q.}\ \bibnamefont
  {He}}\ and\ \bibinfo {author} {\bibfnamefont {Z.-Y.}\ \bibnamefont {Lu}},\
  }\bibfield  {title} {\bibinfo {title} {Quantum renormalization groups based
  on natural orbitals},\ }\href {https://doi.org/10.1103/PhysRevB.89.085108}
  {\bibfield  {journal} {\bibinfo  {journal} {Phys. Rev. B}\ }\textbf {\bibinfo
  {volume} {89}},\ \bibinfo {pages} {085108} (\bibinfo {year}
  {2014})}\BibitemShut {NoStop}%
\bibitem [{\citenamefont {Lu}\ \emph {et~al.}(2014)\citenamefont {Lu},
  \citenamefont {H\"oppner}, \citenamefont {Gunnarsson},\ and\ \citenamefont
  {Haverkort}}]{Lu2014}%
  \BibitemOpen
  \bibfield  {author} {\bibinfo {author} {\bibfnamefont {Y.}~\bibnamefont
  {Lu}}, \bibinfo {author} {\bibfnamefont {M.}~\bibnamefont {H\"oppner}},
  \bibinfo {author} {\bibfnamefont {O.}~\bibnamefont {Gunnarsson}},\ and\
  \bibinfo {author} {\bibfnamefont {M.~W.}\ \bibnamefont {Haverkort}},\
  }\bibfield  {title} {\bibinfo {title} {Efficient real-frequency solver for
  dynamical mean-field theory},\ }\href
  {https://doi.org/10.1103/PhysRevB.90.085102} {\bibfield  {journal} {\bibinfo
  {journal} {Phys. Rev. B}\ }\textbf {\bibinfo {volume} {90}},\ \bibinfo
  {pages} {085102} (\bibinfo {year} {2014})}\BibitemShut {NoStop}%
\bibitem [{\citenamefont {Fishman}\ and\ \citenamefont
  {White}(2015)}]{White2015}%
  \BibitemOpen
  \bibfield  {author} {\bibinfo {author} {\bibfnamefont {M.~T.}\ \bibnamefont
  {Fishman}}\ and\ \bibinfo {author} {\bibfnamefont {S.~R.}\ \bibnamefont
  {White}},\ }\bibfield  {title} {\bibinfo {title} {Compression of correlation
  matrices and an efficient method for forming matrix product states of
  fermionic gaussian states},\ }\href
  {https://doi.org/10.1103/PhysRevB.92.075132} {\bibfield  {journal} {\bibinfo
  {journal} {Phys. Rev. B}\ }\textbf {\bibinfo {volume} {92}},\ \bibinfo
  {pages} {075132} (\bibinfo {year} {2015})}\BibitemShut {NoStop}%
\bibitem [{\citenamefont {Siegbahn}\ \emph {et~al.}(1981)\citenamefont
  {Siegbahn}, \citenamefont {Almlöf}, \citenamefont {Heiberg},\ and\
  \citenamefont {Roos}}]{CASSCF}%
  \BibitemOpen
  \bibfield  {author} {\bibinfo {author} {\bibfnamefont {P.~E.~M.}\
  \bibnamefont {Siegbahn}}, \bibinfo {author} {\bibfnamefont {J.}~\bibnamefont
  {Almlöf}}, \bibinfo {author} {\bibfnamefont {A.}~\bibnamefont {Heiberg}},\
  and\ \bibinfo {author} {\bibfnamefont {B.~O.}\ \bibnamefont {Roos}},\
  }\bibfield  {title} {\bibinfo {title} {The complete active space scf (casscf)
  method in a newton–raphson formulation with application to the hno
  molecule},\ }\href {https://doi.org/10.1063/1.441359} {\bibfield  {journal}
  {\bibinfo  {journal} {The Journal of Chemical Physics}\ }\textbf {\bibinfo
  {volume} {74}},\ \bibinfo {pages} {2384} (\bibinfo {year} {1981})},\ \Eprint
  {https://arxiv.org/abs/https://doi.org/10.1063/1.441359}
  {https://doi.org/10.1063/1.441359} \BibitemShut {NoStop}%
\bibitem [{\citenamefont {Vigman}\ and\ \citenamefont
  {Finkel'shtein}(1978)}]{Vigman1978}%
  \BibitemOpen
  \bibfield  {author} {\bibinfo {author} {\bibfnamefont {P.~B.}\ \bibnamefont
  {Vigman}}\ and\ \bibinfo {author} {\bibfnamefont {A.~M.}\ \bibnamefont
  {Finkel'shtein}},\ }\bibfield  {title} {\bibinfo {title} {Resonant-level
  model in the {K}ondo problem},\ }\href
  {http://www.jetp.ac.ru/cgi-bin/e/index/e/48/1/p102?a=list} {\bibfield
  {journal} {\bibinfo  {journal} {Sov. Phys. JETP}\ }\textbf {\bibinfo {volume}
  {48}},\ \bibinfo {pages} {102} (\bibinfo {year} {1978})}\BibitemShut
  {NoStop}%
\bibitem [{\citenamefont {Giamarchi}(2004)}]{GiamarchiBook}%
  \BibitemOpen
  \bibfield  {author} {\bibinfo {author} {\bibfnamefont {T.}~\bibnamefont
  {Giamarchi}},\ }\href@noop {} {\emph {\bibinfo {title} {Quantum Physics in
  One Dimension}}}\ (\bibinfo  {publisher} {Oxford University Press},\ \bibinfo
  {address} {New York},\ \bibinfo {year} {2004})\BibitemShut {NoStop}%
\bibitem [{\citenamefont {Gogolin}\ \emph {et~al.}(2004)\citenamefont
  {Gogolin}, \citenamefont {Nersesyan},\ and\ \citenamefont
  {Tsvelik}}]{GogolinBook}%
  \BibitemOpen
  \bibfield  {author} {\bibinfo {author} {\bibfnamefont {A.~O.}\ \bibnamefont
  {Gogolin}}, \bibinfo {author} {\bibfnamefont {A.~A.}\ \bibnamefont
  {Nersesyan}},\ and\ \bibinfo {author} {\bibfnamefont {A.~M.}\ \bibnamefont
  {Tsvelik}},\ }\bibinfo {title} {{Bosonization and Strongly Correlated
  Systems}}\ (\bibinfo  {publisher} {Cambridge University Press},\ \bibinfo
  {year} {2004})\BibitemShut {NoStop}%
\bibitem [{\citenamefont {Weiss}(2012)}]{WeissBook}%
  \BibitemOpen
  \bibfield  {author} {\bibinfo {author} {\bibfnamefont {U.}~\bibnamefont
  {Weiss}},\ }\href {https://doi.org/10.1142/8334} {\emph {\bibinfo {title}
  {Quantum Dissipative Systems}}},\ \bibinfo {edition} {4th}\ ed.\ (\bibinfo
  {publisher} {WORLD SCIENTIFIC},\ \bibinfo {year} {2012})\ \Eprint
  {https://arxiv.org/abs/https://www.worldscientific.com/doi/pdf/10.1142/8334}
  {https://www.worldscientific.com/doi/pdf/10.1142/8334} \BibitemShut {NoStop}%
\bibitem [{\citenamefont {Guinea}\ \emph {et~al.}(1985)\citenamefont {Guinea},
  \citenamefont {Hakim},\ and\ \citenamefont {Muramatsu}}]{Guinea}%
  \BibitemOpen
  \bibfield  {author} {\bibinfo {author} {\bibfnamefont {F.}~\bibnamefont
  {Guinea}}, \bibinfo {author} {\bibfnamefont {V.}~\bibnamefont {Hakim}},\ and\
  \bibinfo {author} {\bibfnamefont {A.}~\bibnamefont {Muramatsu}},\ }\bibfield
  {title} {\bibinfo {title} {Bosonization of a two-level system with
  dissipation},\ }\href {https://doi.org/10.1103/PhysRevB.32.4410} {\bibfield
  {journal} {\bibinfo  {journal} {Phys. Rev. B}\ }\textbf {\bibinfo {volume}
  {32}},\ \bibinfo {pages} {4410} (\bibinfo {year} {1985})}\BibitemShut
  {NoStop}%
\bibitem [{\citenamefont {Zar\'and}\ and\ \citenamefont {von
  Delft}(2000)}]{Zarand00}%
  \BibitemOpen
  \bibfield  {author} {\bibinfo {author} {\bibfnamefont {G.}~\bibnamefont
  {Zar\'and}}\ and\ \bibinfo {author} {\bibfnamefont {J.}~\bibnamefont {von
  Delft}},\ }\bibfield  {title} {\bibinfo {title} {Analytical calculation of
  the finite-size crossover spectrum of the anisotropic two-channel kondo
  model},\ }\href {https://doi.org/10.1103/PhysRevB.61.6918} {\bibfield
  {journal} {\bibinfo  {journal} {Phys. Rev. B}\ }\textbf {\bibinfo {volume}
  {61}},\ \bibinfo {pages} {6918} (\bibinfo {year} {2000})}\BibitemShut
  {NoStop}%
\bibitem [{\citenamefont {Nghiem}\ \emph {et~al.}(2016)\citenamefont {Nghiem},
  \citenamefont {Kennes}, \citenamefont {Kl\"ockner}, \citenamefont {Meden},\
  and\ \citenamefont {Costi}}]{Nghiem2016}%
  \BibitemOpen
  \bibfield  {author} {\bibinfo {author} {\bibfnamefont {H.~T.~M.}\
  \bibnamefont {Nghiem}}, \bibinfo {author} {\bibfnamefont {D.~M.}\
  \bibnamefont {Kennes}}, \bibinfo {author} {\bibfnamefont {C.}~\bibnamefont
  {Kl\"ockner}}, \bibinfo {author} {\bibfnamefont {V.}~\bibnamefont {Meden}},\
  and\ \bibinfo {author} {\bibfnamefont {T.~A.}\ \bibnamefont {Costi}},\
  }\bibfield  {title} {\bibinfo {title} {Ohmic two-state system from the
  perspective of the interacting resonant level model: Thermodynamics and
  transient dynamics},\ }\href {https://doi.org/10.1103/PhysRevB.93.165130}
  {\bibfield  {journal} {\bibinfo  {journal} {Phys. Rev. B}\ }\textbf {\bibinfo
  {volume} {93}},\ \bibinfo {pages} {165130} (\bibinfo {year}
  {2016})}\BibitemShut {NoStop}%
\bibitem [{\citenamefont {Blunden-Codd}\ \emph {et~al.}(2017)\citenamefont
  {Blunden-Codd}, \citenamefont {Bera}, \citenamefont {Bruognolo},
  \citenamefont {Linden}, \citenamefont {Chin}, \citenamefont {von Delft},
  \citenamefont {Nazir},\ and\ \citenamefont {Florens}}]{BeraSubohmic}%
  \BibitemOpen
  \bibfield  {author} {\bibinfo {author} {\bibfnamefont {Z.}~\bibnamefont
  {Blunden-Codd}}, \bibinfo {author} {\bibfnamefont {S.}~\bibnamefont {Bera}},
  \bibinfo {author} {\bibfnamefont {B.}~\bibnamefont {Bruognolo}}, \bibinfo
  {author} {\bibfnamefont {N.-O.}\ \bibnamefont {Linden}}, \bibinfo {author}
  {\bibfnamefont {A.~W.}\ \bibnamefont {Chin}}, \bibinfo {author}
  {\bibfnamefont {J.}~\bibnamefont {von Delft}}, \bibinfo {author}
  {\bibfnamefont {A.}~\bibnamefont {Nazir}},\ and\ \bibinfo {author}
  {\bibfnamefont {S.}~\bibnamefont {Florens}},\ }\bibfield  {title} {\bibinfo
  {title} {Anatomy of quantum critical wave functions in dissipative impurity
  problems},\ }\href {https://doi.org/10.1103/PhysRevB.95.085104} {\bibfield
  {journal} {\bibinfo  {journal} {Phys. Rev. B}\ }\textbf {\bibinfo {volume}
  {95}},\ \bibinfo {pages} {085104} (\bibinfo {year} {2017})}\BibitemShut
  {NoStop}%
\bibitem [{\citenamefont {Hanl}\ and\ \citenamefont
  {Weichselbaum}(2014)}]{Weichselbaum2014}%
  \BibitemOpen
  \bibfield  {author} {\bibinfo {author} {\bibfnamefont {M.}~\bibnamefont
  {Hanl}}\ and\ \bibinfo {author} {\bibfnamefont {A.}~\bibnamefont
  {Weichselbaum}},\ }\bibfield  {title} {\bibinfo {title} {Local susceptibility
  and kondo scaling in the presence of finite bandwidth},\ }\href
  {https://doi.org/10.1103/PhysRevB.89.075130} {\bibfield  {journal} {\bibinfo
  {journal} {Phys. Rev. B}\ }\textbf {\bibinfo {volume} {89}},\ \bibinfo
  {pages} {075130} (\bibinfo {year} {2014})}\BibitemShut {NoStop}%
\end{thebibliography}
\end{document}